\documentclass{emulateapj}

\shorttitle{A Global Stability Analysis of Galaxy Clusters}
\shortauthors{Guo et al.}

\begin{document}

\title{A Global Stability Analysis of Clusters of Galaxies with Conduction and AGN Feedback Heating} 

\author{Fulai Guo\altaffilmark{1}, S. Peng Oh \altaffilmark{1} and M. Ruszkowski\altaffilmark{2}}

\altaffiltext{1}{Department of Physics, University of California, Santa Barbara, CA 93106, USA;\\
E-mail: johnnie@physics.ucsb.edu (FG); peng@physics.ucsb.edu (SPO)}
\altaffiltext{2}{Department of Astronomy, The University of Michigan, 500 Church Street, Ann Arbor, MI 48109, USA;\\
E-mail: mateuszr@umich.edu}

\begin{abstract}
We investigate a series of steady-state models of galaxy clusters, in which the hot intracluster gas is efficiently heated by active galactic nucleus (AGN) feedback and thermal conduction, and in which the mass accretion rates are highly reduced compared to those predicted by the standard cooling flow models. We perform a global Lagrangian stability analysis. We show for the first time that the global radial instability in cool core clusters can be suppressed by the AGN feedback mechanism, provided that the feedback efficiency exceeds a critical lower limit. 
Furthermore, our analysis naturally shows that the clusters can exist in two distinct forms.
Globally stable clusters are expected to have either: 1) cool cores 
stabilized by both AGN feedback and conduction, or 2) non-cool cores stabilized primarily by conduction. Intermediate central temperatures typically lead to globally unstable solutions. This bimodality is consistent with the recently observed anticorrelation between the flatness of the temperature profiles and the AGN activity (Dunn \& Fabian 2008)
and the observation by Rafferty et al. (2008) that the shorter central cooling times tend to correspond to 
significantly younger AGN X-ray cavities.
\end{abstract}


\section{Introduction}
\label{section:intro}

Clusters of galaxies are the largest gravitationally bound systems in the universe. They are filled with hot gas with $T\sim 2-10$ keV, which loses thermal energy prolifically by emitting X-rays. The X-ray surface brightness of many galaxy clusters shows a strong central peak that was previously interpreted as the signature of a cooling flow with mass accretion rates of up to several hundred M$_{\sun}$ yr$^{-1}$ \citep[see][for a review]{1994ARA&A..32..277F}. Although the gas temperature is observed to decline toward cluster centers, recent high-resolution \textit{Chandra} and \textit{XMM-Newton} observations show a remarkable lack of emission lines from the gas at temperatures below about $\sim 1/3$ of the ambient cluster temperature (e.g., \citealt{2001A&A...365L.104P, 2003ApJ...590..207P, 2001A&A...365L..87T}; for a review see \citealt{2006PhR...427....1P}). In addition, the spectroscopically determined mass deposition rates are significantly smaller than the classic values estimated from the X-ray luminosity within the cooling regions \citep{2004MNRAS.347.1130V}. The absence of a cool phase in cores of galaxy clusters is suggestive of one or more heating mechanisms maintaining the hot gas at keV temperatures for a period at least comparable to the lifetime of galaxy clusters.

Amongst the many candidate heating mechanisms put forth recently, there are two leading contenders: 
\begin{enumerate}
  \item thermal conduction from the hot outer regions of the cluster to the center (e.g., \citealt{1986ApJ...306L...1B}; \citealt{2003ApJ...582..162Z}, hereafter ZN03; \citealt{2004MNRAS.347.1130V}); 
  \item heating of the intracluster medium (ICM) by outflows, bubbles, or cosmic rays generated by AGNs at cluster centers (e.g., \citealt{2002Natur.418..301B}, \citealt{2002ApJ...581..223R}, \citealt{2004ApJ...611..158R}, \citealt{2007ApJ...671.1413C}, \citealt{2007arXiv0706.1274G}, Sijacki et al. 2008).
\end{enumerate}

Recent theoretical and numerical work \citep[e.g.][]{2001ApJ...562L.129N, 2003ApJ...589L..77C} has shown that a turbulent magnetic field is not as efficient in suppressing thermal conduction as previously thought. In particular, \citet{2001ApJ...562L.129N} showed that the effective thermal conductivity $\kappa$ in a turbulent MHD medium is a substantial fraction ($\sim 1/5$) of the classical Spitzer value $\kappa_{\rm{Sp}}$ if magnetic turbulence extends over at least two decades in scale. On the other hand, recent work shows that bouyancy instabilities could potentially strongly suppress conductivity in the cluster core \citep{eliot08,eliot_parrish08}, a point we discuss in \S\ref{section:conclusion}. Following this work, ZN03 shows that the electron density and temperature profiles of half of the clusters they investigated can be fitted by a pure conduction model with the conductivity suppression factor $f \equiv \kappa/\kappa_{\rm{Sp}}\sim 0.2-0.4$. However, if only thermal conduction operates to balance the cooling, extreme fine-tuning of the conduction suppression factor $f$ is required (\citealt{2007arXiv0706.1274G}; also see \citealt{1988ApJ...326..639B}): if $f$ is too low, then a strong cooling flow develops, while if $f$ is too high, the temperature profile becomes nearly isothermal, in contrast to observations of cool core clusters where the temperature invariably declines toward the cluster center. Furthermore, although thermal conduction is well known to stabilize short-wavelength perturbations against thermal instability (e.g., \citealt{1965ApJ...142..531F}; \citealt{1987ApJ...319..632M}), the pure conduction models of the cool core clusters are thermally unstable against global perturbations (\citealt{2003MNRAS.342..463S}; \citealt{2003ApJ...596..889K}, hereafter KN03). Using a Lagrangian perturbation analysis, KN03 showed that the pure conduction model has one globally unstable radial mode with the instability growth  (e-folding) time of $\sim 2-5$ Gyr. Furthermore, if strong perturbations are applied (as would be the case in, for instance, a cluster merger), the growth times can be even shorter. \citet{2007arXiv0706.1274G} showed that if one started from arbitrary initial conditions (rather than an equilibrium solution), a catastrophic cooling flow quickly develops in a conduction-only model with a moderate level of conductivity. 

Fortunately, other sources of heating exist. A particularly promising candidate is heating by the central AGN, for which observational evidence has been growing in recent years (see \citealt{2007ARA&A..45..117M} for a recent review). A majority ($\sim 71\%$) of cool core clusters harbor radio sources at their cluster centers \citep{1990AJ.....99...14B}. Following the launch of  \textit{Chandra} and \textit{XMM-Newton}, recent high-resolution X-ray observations also indicate that these radio sources are interacting with their surroundings and often displace the ICM, producing X-ray cavities (e.g., \citealt{2000MNRAS.318L..65F}, \citealt{2004ApJ...607..800B}, \citealt{2007ApJ...665.1057F}). 

By contrast with conduction, heating by the dissipation of mechanical energy released by central AGNs provides a self-regulating feedback mechanism (e.g., \citealt{2001ApJ...551..131C}, \citealt{2002ApJ...581..223R}, \citealt{2003ApJ...587..580B}, \citealt{2003MNRAS.338..837K}, \citealt{2007arXiv0706.1274G}). If AGN activity is triggered by cooling-induced gas accretion toward cluster centers, the AGN heating increases until it halts further accretion. Thus, the gas accretion rate is self-regulated as brief bursts of AGN activity alternate with cooling (e.g., \citealt{2005ApJ...634..955V}). Due to this AGN feedback heating, the accretion flow may automatically adjust itself to a low value of the accretion rate, which depends mainly on the feedback efficiency $\epsilon$ (see equation \ref{depsilon}), and, in a time-averaged sense, the ICM may reach a quasi-equilibrium state. This has been clearly demonstrated by \citet[][hereafter RB02]{2002ApJ...581..223R} in hydrodynamic simulations, who showed that a model cluster heated by a combination of thermal conduction and AGN feedback does not suffer from the cooling catastrophe, but instead relaxes to a stable quasi-equilibrium state. More recently, \citet{2007arXiv0706.1274G} proposed a new model of AGN feedback heating, where the ICM is efficiently heated by both thermal conduction and the cosmic rays produced by accretion-triggered AGN activity. In their model, the ICM also relaxes to a stable steady state with the mass accretion rate highly reduced, and, more importantly, their results do not require fine tuning of the various adjustable parameters, including thermal conductivity and the AGN heating efficiency. Moreover, unlike the conduction-only case, the simulation relaxes to a stable state independent of the initial conditions. Although the detailed dynamics of how the released AGN energy is transferred into thermal energy of the ICM may be much more complicated than these models and is still poorly understood at the present time, these simulations strongly indicate that AGN feedback heating may potentially solve the fine-tuning problem associated with the pure conduction model. 

These simulations also suggest that AGN feedback heating plays a key role in suppressing global thermal instability in the ICM (see \citealt{1989ApJ...338..761R} for a local analysis). While local thermal instability may only produce small-scale structures (e.g., local mass dropout, emission-line filaments) in galaxy clusters, global thermal instability may result in a cooling catastrophe and a strong cooling flow. Thus, a successful model for the ICM must be globally stable, or at least only have instabilities which grow on extremely long timescales. In the present paper, we will use the Lagrangian perturbation method to formally investigate thermal instability in quasi-equilibrium galaxy clusters with thermal conduction and AGN feedback heating. Global stability analysis is a method complementary to numerical simulations as it allows for the quick identification of global trends, quick systematic parameter search and helps to build physical intuition.
For the spatial distribution of AGN feedback heating, we adopt the analytically tractable model proposed by \citet{2001ASPC..240..363B}. This model has been compared with observations by \citet{2006A&A...453..423P}, who show that the model usually provides a satisfactory explanation of the observed structure of cool core clusters, although in a fair fraction of their sample the model provide relatively poor fits. However, our emphasis is not on the background solutions of this particular model but their global stability. With slight modification, our methods can be applied to {\it any} model of AGN feedback heating.  

We will show that the feedback mechanism can indeed effectively suppress global radial thermal instability in cool core (CC) clusters, provided that the AGN feedback efficiency is larger than a lower limit (see \S~\ref{section:globalstability} and \S~\ref{section:deppara} for details). We will also study the dependence of the cluster stability on the background ICM profiles (\S~\ref{section:depsteady}): for non-cool core (NCC) clusters, which have relatively flat temperature profiles and which are less studied in the literature, the stabilizing effect of the feedback mechanism becomes small, but thermal conduction may completely suppress global thermal instability. Thus, we propose that thermal stability of the ICM favors two distinct categories of cluster steady state profiles: CC clusters stabilized mainly by AGN feedback and NCC clusters stabilized by thermal conduction. Interestingly, X-ray observations suggest that clusters can be subdivided into two distinct categories according to the presence or absence of a cool core (e.g., Peres et al. 1998, Bauer et al. 2005, Sanderson et al. 2006, Chen et al. 2007). Our stability analysis thus naturally explains these two distinct cluster categories. Our model is also consistent with the observation by Rafferty et al. (2008) who show
that the short central cooling time corresonds to younger AGN (i.e., shorter X-ray cavity ages)
and the anticorrelation between the flatness of the temperature profiles and the AGN activity recently reported by Dunn \& Fabian 2008.

The issue of the formation and evolution of NCC and CC clusters has recently been addressed by Burns et al. (2008)
who performed large scale cosmological simulations. They found out that the CC and NCC clusters follow different evolutionary tracks,
with CC clusters accreting more slowly over time and growing enhanced cool cores via hierarchical mergers.  In contrast, they argued that
NCC suffered early mergers that disrupted embryonic cool cores. However, this pioneering work does not include the effects of
AGN to stop catastrophic cooling in the centers of CCs and the numerical resolution in their simulations is still 
too low (15.6$h^{-1}$ kpc) to accurately study the stability and structure of the cores once they are formed.
McCarthy et al. (2008) invoked different preheating histories to explain the difference between CC and NCC clusters. Like us, they find that AGN heating is required to stabilize CC clusters. Our present calculations make the new suggestion that the bifircation between CC and NCC clusters emerges naturally, from the fact that clusters with intermediate central temperatures are globally unstable. 

The rest of the paper is organized as follows. In \S~\ref{section:steady}, we describe the time-dependent equations of the thermal intracluster gas and construct a series of steady-state cluster models, in which the mass accretion rate is highly suppressed compared to that predicted by standard cooling flow models. We then carry out a detailed formal linear stability analysis of local and global modes of thermal instability in steady-state galaxy clusters in \S~\ref{section:stability}, where we also study the dependence of the cluster stability on the AGN feedback efficiency and on the background cluster profiles. We summarize our main results in \S~\ref{section:conclusion} with a discussion of the implications. The cosmological parameters used throughout this paper are: $\Omega_{m}=0.3$, $\Omega_{\Lambda}=0.7$, $h=0.7$. We have rescaled observational results if the original paper used a different cosmology.

\section{Steady-State Models}
\label{section:steady}

\subsection{Time-dependent equations}

In our model, the intracluster medium is subject to radiative cooling, AGN feedback heating and thermal conduction. The governing hydrodynamic equations are

\begin{eqnarray}
\frac{d \rho}{d t} + \rho \nabla \cdot {\bf v} = 0,\label{hydro1}
\end{eqnarray}
\begin{eqnarray}
\rho \frac{d {\bf v}}{d t} = -\nabla P-\rho \nabla \Phi ,\label{hydro2}
\end{eqnarray}
\begin{eqnarray}
\frac{1}{\gamma-1}\frac{d P}{d t} -\frac{\gamma}{\gamma-1}\frac{ P}{\rho} \frac{d \rho}{d t}= \mathcal{H} - \nabla \cdot {\bf F}-\rho \mathcal{L}
   ,\label{hydro3}\\ \nonumber
\end{eqnarray}

\noindent
where $d/dt \equiv \partial/\partial t+{\bf v}  \cdot \nabla $ is the Lagrangian time derivative,
$\rho$ is the gas density, $P$ is the gas pressure, ${\bf v}$ is the gas velocity, $\Phi$ is the gravitational potential, $\gamma=5/3$ is the adiabatic index of thermal gas, $\rho \mathcal{L}=n_{e}^{2}\Lambda(T)=2.1 \times 10^{-27}n_{e}^{2}T^{1/2}$ ergs cm$^{-3}$ s$^{-1}$ is the volume cooling rate due to thermal bremsstrahlung\footnote{We have simplified the form of the cooling function here, ignoring the contribution of metal line cooling, which is important at lower temperatures. Fits to the full cooling function do exist \citep{1993ApJS...88..253S}, but they complicate the analytic derivation of the global stability analysis. 
We have experimented with the full cooling function and didn't find qualitative changes to our results (see Fig. \ref{plot11}).} (\citealt{1979rpa..book.....R}; KN03), and ${\bf F}$ is the conductive heat flux  

\begin{eqnarray}
  {\bf F} = -\kappa \nabla T , \label{conduction}\\ \nonumber
\end{eqnarray}

\noindent
where $\kappa$ is the effective isotropic conductivity. Depending on the details of plasma magnetization and MHD turbulence, heat transport in the ICM may be very complex, and both electron conduction and turbulent mixing may contribute to heat transport (e.g., \citealt{2006ApJ...645L..25L}); additionally various instabilities could alter the nature of conductivity within the cooling region \citep{eliot08,eliot_parrish08}. Since at present there is no consensus on the nature of conductivity in a turbulent magnetized plasma, we adopt the same assumption of Spitzer conductivity (with a factor $f$ due to magnetic field suppression) that most authors do (e.g., ZN03, KN03), 

\begin{eqnarray}
\kappa =f\kappa_{\rm{Sp}}  , \label{conductivity}\\ \nonumber
\end{eqnarray}

\noindent
where $\kappa_{Sp}$ is the classical Spitzer conductivity \citep*{1962pfig.book.....S},
 \begin{eqnarray}
  \kappa_{\rm{Sp}}= \frac{1.84 \times 10^{-5}}{\ln \lambda }T^{5/2} 
\rm{ ergs}\;\rm{s}^{-1}\rm{K}^{-7/2}\rm{ cm} ^{-1},\label{spitzer}\\ \nonumber
\end{eqnarray}

\noindent
with the usual Coulomb logarithm $\rm{ln} \lambda \sim 37$. In this paper, we assume that $f$ ($0\leq f \leq 1$) is constant in both space and time.

According to the ideal gas law, the gas pressure is related to the gas temperature $T$ and the electron number density $n_{e}$ via

\begin{eqnarray}
 P =\frac{\rho k_{B} T}{\mu m_{\mu}}=\frac{\mu _{e}}{\mu}n_{e}k_{B}T , \label{estate}\\ \nonumber
\end{eqnarray}

\noindent
where $k_{B}$ is Boltzmann's constant, $m_{\mu}$ is the atomic mass unit, and $\mu$ and $\mu_{e}$ are the mean molecular weight per thermal particle and per electron, respectively. 
As in ZN03, we use $\mu=0.62$ and $\mu_{e}=1.18$, corresponding to a fully ionized gas with hydrogen fraction $X=0.7$ and helium fraction $Y=0.28$.

In equation~(\ref{hydro2}), we neglect the self-gravity of the gas and any dynamical effects of magnetic fields. The gravitational potential $\Phi$ is determined by the dark matter distribution $\rho_{\rm DM}$, which we assume has a modified Navarro-Frenk-White (NFW) form \citep{1997ApJ...490..493N} with a softened core (ZN03):

\begin{eqnarray}
\rho_{\rm{DM}}(r)=\frac{M_{0}/2\pi}{(r+r_{c})(r+r_{s})^{2}} ,\\ \nonumber  
\end{eqnarray}

\noindent
where $r_{c}$ is a softening radius which introduces a core in the dark matter distribution in the very inner regions, $r_{s}$ is  the standard scale radius  of the NFW profile and $M_{0}$ is a characteristic mass. The corresponding gravitational potential is:

\begin{eqnarray}
\Phi & = & - 2GM_{0}  \frac{r_{c}}{(r_{s}-r_{c})^{2}}\left [ \ln\frac{1+r/r_{c}}{1+r/r_{s}}   + \frac{\ln(1+r/r_{c})}{r/r_{c}}\right] \nonumber \\
     &   & - 2GM_{0}\frac{r_{s}(r_{s}-2r_{c}) }{r_{c}(r_{s}-r_{c})^{2}}\frac{\ln (1+r/r_{s})}{r/r_{c}}  ,\\ \nonumber 
\end{eqnarray}

\noindent
where $G$ is the gravitational constant. The parameters $M_{0}$ and $r_{s}$ for a given cluster are obtained from the observed temperature in the outer regions of the cluster, as described in ZN03. We adopt their calculated values of these two parameters and their best-fit values of $r_{c}$ directly.

The term $\mathcal{H}$ in equation~(\ref{hydro3}) is the volume heating rate due to AGN feedback.
We adopt the ``effervescent heating" mechanism proposed by \citet{2001ASPC..240..363B} to describe 
the energy deposition into the ICM by the rising bubbles, which are produced by the central AGN.
Because of the non-negligible gas pressure gradient in the ICM, the bubbles will expand as they rise, doing $pdV$ work and converting the internal bubble energy to kinetic energy of the ICM. The resulting disorganized motion of the ICM is quickly converted to heat. Assuming that this heating mechanism reaches a quasi-steady state, the details of the bubble filling factor, rise rate and geometry should cancel. If the cavity expands adiabatically (see \citet{2007arXiv0706.1274G} for an alternative scenario), the luminosity passing through the surface of a sphere at radius $r$ is:

\begin{eqnarray}
{\bf \dot{E}} \propto p_{b}(r)^{(\gamma_{b}-1)/\gamma_{b}} {\bf\hat{r}} ,  \label{agnflux}\\ \nonumber
\end{eqnarray}

\noindent
where $p_{b}(r)$ is the pressure of buoyant gas inside bubbles, $\gamma_{b} \approx 4/3$ is the adiabatic index of buoyant gas (assuming it is primarily composed of relativisitic plasma), and ${\bf\hat{r}}$ is the unit vector along the radial direction. Assuming that the bubble rises subsonically so that pressure equilibrium is maintained, $p_{b}(r)=P(r)$, where $P(r)$ is the thermal pressure of the ICM, we may rewrite equation (\ref{agnflux}) as:

\begin{eqnarray}
{\bf\dot{E}} \sim L_{\rm{agn}}\left(\frac{P}{P_{0}}\right)^{\beta} {\bf\hat{r}} ,  \label{agnheat0}\\ \nonumber
\end{eqnarray}

\noindent
where $P_{0}$ is the gas pressure at the cluster center, $L_{\rm{agn}}$ is the AGN mechanical luminosity, and $\beta=(\gamma_{b}-1)/\gamma_{b}$.

AGN activity is likely to be intermittent on a timescale of order the Salpeter time $t_{\rm S} \sim 10^{7}$ yr, and possibly as short as $t_{i} \sim 10^{4}-10^{5}$ yr (\citealt{1997ApJ...487L.135R}). Note that the bubble rise time is  typically comparable to (at most several times) the sound crossing time $t_{\rm sc} \sim 10^{8} r_{100} c_{s,1000}^{-1}$yr for a radius $r \sim 100 r_{100}$ kpc and sound speed $c_{\rm s} \sim 1000 c_{s,1000} \, {\rm km \, s^{-1}}$ (e.g., see table 3 in \citet{2004ApJ...607..800B}), and is usually shorter than the gas cooling time. Thus, it is justifiable to treat AGN heating in a time-averaged sense and assume that the mechanical energy of central AGN is injected into the whole ICM instantaneously (e.g., RB02; \citealt{2003ApJ...587..580B}). We further assume the AGN mechanical luminosity to be 

\begin{eqnarray}
L_{\rm{agn}} = -\epsilon \dot{M}_{\rm{in}}c^{2} ,  
\label{depsilon}\\ \nonumber
\end{eqnarray}

\noindent
where $\epsilon$ is the kinetic efficiency of AGN feedback, and $\dot{M}_{\rm{in}}=4\pi r_{\rm{in}}^{2}\rho_{0} v_{0}$ is the mass accretion rate at the inner radius $r_{\rm{in}}$, where $\rho_{0}$ and $v_{0}$ are the density and radial velocity of thermal gas at $r_{\rm{in}}$, respectively. Therefore, the volume AGN heating rate $\mathcal{H}$ may be written as

\begin{eqnarray}
\mathcal{H} & \sim & - \nabla \cdot \frac{{\bf\dot{E}}}{4\pi r^{2}}  \nonumber \\
            & \sim &  \frac{\epsilon \beta \dot{M}_{\rm{in}}c^{2}}{4\pi r^{3}} \left( 1-e^{-r/r_{0}}\right)\left(\frac{P}{P_{0}}\right)^{\beta}\frac{\partial \ln P}{\partial\ln r} ,
\label{agnheat}\\ \nonumber
\end{eqnarray}

\noindent
where $r_{0}$ is the inner heating cutoff radius, which is determined by the finite size of the central radio source (\citealt{2002ApJ...581..223R}; also see a discussion of $r_{0}$ in \citealt{2005ApJ...634...90R}). In the rest of this paper, $r_{0}$ is taken to be $20$ kpc, unless otherwise stated.

\subsection{Steady-state models}
\label{section:stst}

\begin{table*}
 \centering
 \begin{minipage}{160mm}
  \renewcommand{\thefootnote}{\thempfootnote} 
  \caption{Parameters and results of the steady-state models for typical cool core clusters}
  \vspace{3mm}
  \begin{tabular}{@{}llllcccccc}
  \hline
  \hline
        & {$T_{\rm{in}}$\footnote{
These boundary values are adopted from the best-fit models of ZN03, unless otherwise stated.}}
& {$T_{\rm{out}}$\footnotemark[\it{a}]} & {$n_{0}$\footnotemark[\it{a}]}   & & $\dot{M}$ &   & & {$n_{\rm{out}}$\footnote{$n_{\rm{out}}$ is the corresponding model electron number density at the outer boundary $r_{\rm{out}}$.}}&\\
     Name &(keV) &(keV)&(cm$^{-3})$ &Model&($M_{\sun}/$yr) &$\epsilon$ & $f$ &(cm$^{-3})$&{$h_{\rm{agn}}/L_{X}$ 
     \footnote{$h_{\rm{agn}}/L_{X}$ is the ratio of the overall AGN heating rate to X-ray luminosity of the cluster.}}   \\
 \hline
 A1795 & 2 & 7.5 & {0.053 \footnote{ Adopted from the \textit{Chandra} observation \citep{2002MNRAS.331..635E}.}} &A1&  $-$0.15& 0 & 0.27&$1.42\times10^{-4}$&0 \\
             &&&&A2&$-$0.15&0.1&0.12&$1.26\times10^{-4}$&0.52 \\
             &&&&A3&$-$0.05&0.3&0.12 & $1.26\times10^{-4}$&0.52\\
A2199 & 1.6 & 5 & 0.074 & B1 & $-$0.015  & 0 &0.43&$4.22\times10^{-5}$&0\\
             &&&&B2&$-$0.015&0.05&0.36&$4.00\times10^{-5}$&0.12 \\ 
              &&&&B3&$-$0.00375&0.2&0.36&$4.00\times10^{-5}$&0.12 \\                           
  A2052& 1.3 & 3.5 & {0.035  \footnote{ Adopted from the \textit{Chandra} observation \citep{2001ApJ...558L..15B}.}} & C1 & $-$0.006 & 0&0.31&$1.78\times10^{-5}$&0 \\
               &&&&C2&$-$0.006&0.05&0.20&$1.62\times10^{-5}$&0.29 \\
              &&&&C3&$-$0.0015&0.2&0.20&$1.62\times10^{-5}$&0.29 \\
  {A2597 
  \footnote{To provide a better fit to observations, we take the outer boundary of the cluster A2597 to be $r_{{\rm{out}}}=300$ kpc and the AGN heating cutoff radius to be $r_{0}=40$ kpc.}} 
  & 1  & 4 & {0.07 \footnote{ Adopted from the \textit{Chandra} observation \citep{2001ApJ...562L.149M}.}} &  D1 &  $-$0.32 & 0&1.30&$1.39\times10^{-3}$&0 \\
              &&&&D2&$-$0.32&0.05&0.40&$1.31\times10^{-3}$&0.68 \\ 
              &&&&D3&$-$0.16&0.1&0.40&$1.31\times10^{-3}$&0.68 \\             
              &&&&D4&$-$0.17&0.1&0.30&$1.30\times10^{-3}$&0.76 \\ 
              &&&&D5&$-$0.14&0.1&0.50&$1.33\times10^{-3}$&0.60 \\ 
\hline
\label{table1}
\end{tabular}
\end{minipage}
\end{table*}

\begin{figure*}
\plotone{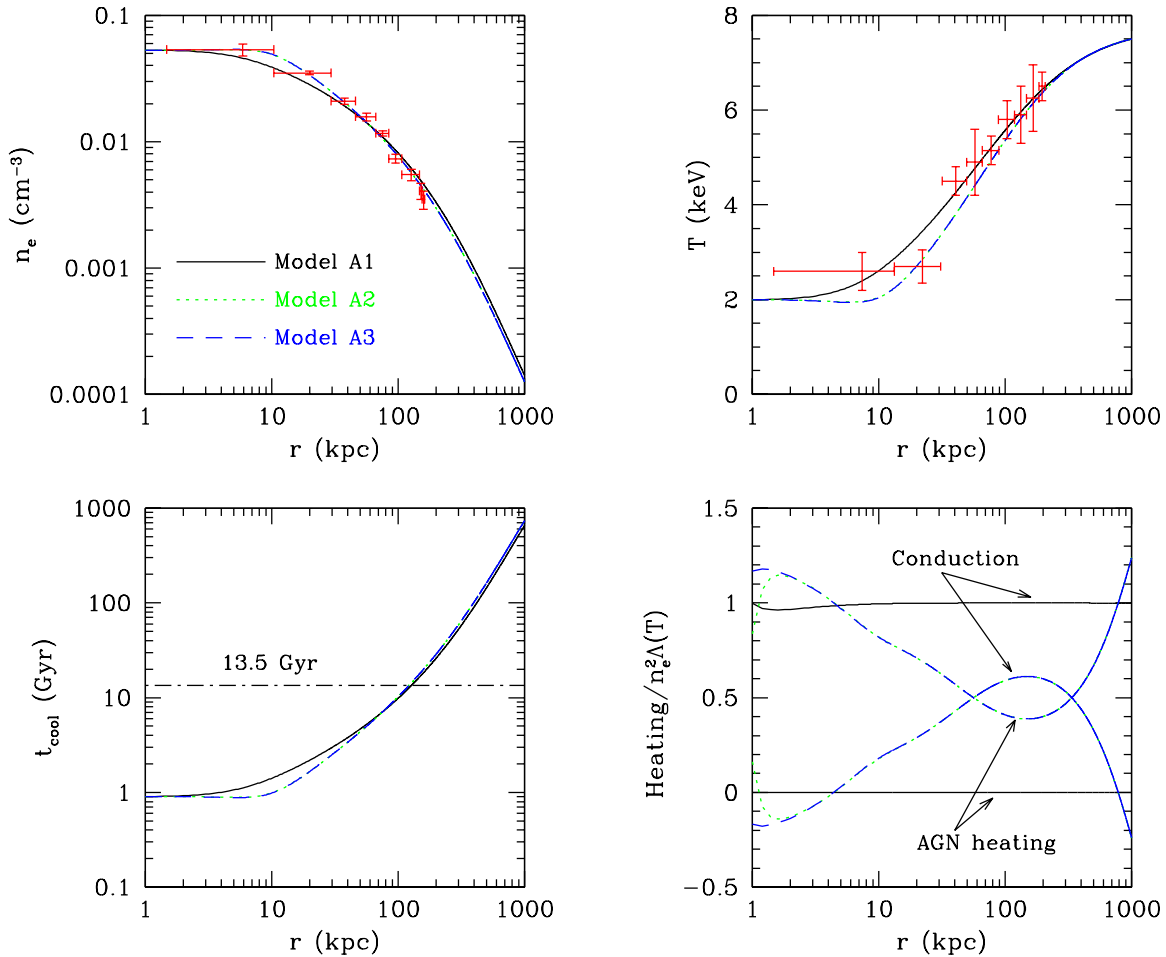}
\caption{Electron number density (\textit{upper left}), temperature (\textit{upper right}), isobaric cooling time (\textit{lower left}), and relative importance of AGN heating and conduction (\textit{lower right}) in three typical steady-state models of the cluster Abell 1795. The dot-dashed line in the \textit{lower left} panel shows the age of the universe ($13.5$ Gyr for the cosmology used in this paper). Crosses in the {upper} panels correspond to \textit{Chandra} data \citep{2002MNRAS.331..635E}. See text and Table \ref{table1} for additional information.}
 \label{plotone}
 \end{figure*}

\begin{figure*}
\plotone{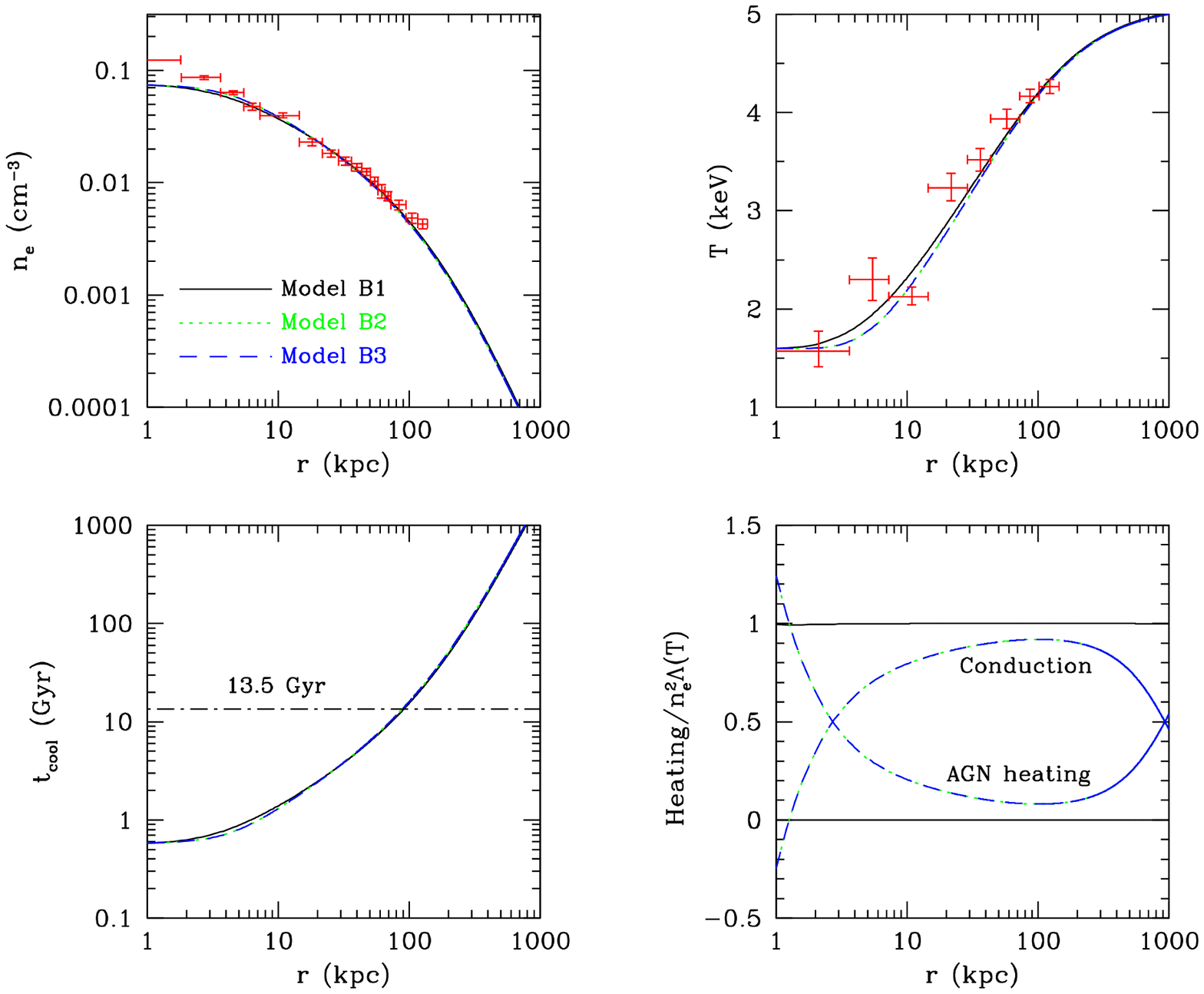}
\caption{Electron number density (\textit{upper left}), temperature (\textit{upper right}), isobaric cooling time (\textit{lower left}), and relative importance of AGN heating and conduction (\textit{lower right}) in three typical steady-state models of the cluster Abell 2199. Crosses in the {upper} panels correspond to \textit{Chandra} data \citep{2002MNRAS.336..299J}. See text and Table \ref{table1} for additional information.}
 \label{plottwo}
 \end{figure*}
 
In this subsection, we will construct steady-state cluster models, which will be used as initial unperturbed states in stability analysis presented in the next section. We assume that the cluster is spherically symmetric and time independent. Equations~(\ref{hydro1}), (\ref{hydro2}) and (\ref{hydro3}) are thus simplified to

\begin{eqnarray}
\dot{M}\equiv 4\pi r^{2}\rho v=\rm{constant} ,\label{hydro4}
\end{eqnarray}
\begin{eqnarray}
\rho v \frac{dv}{d r} = -\frac{dP}{dr}-\rho \frac{d \Phi}{dr} ,\label{hydro5}
\end{eqnarray}
\begin{eqnarray}
\frac{v}{\gamma-1}\frac{d P}{d r} -\frac{\gamma}{\gamma-1}\frac{ Pv}{\rho} \frac{d\rho }{d r}= \mathcal{H} - \frac{1}{r^{2}}\frac{d}{dr}(r^{2}F)-\rho \mathcal{L}
   ,\label{hydro6}\\ \nonumber
\end{eqnarray}

\noindent
where $v$ is the radial gas speed and $F$ is the radial heat flux.

Equations (\ref{conduction}), (\ref{hydro5}) and (\ref{hydro6}) are three ordinary differential equations for the three variables $P(r)$, $T(r)$ and $r^{2}F(r)$. We solve these equations as a boundary value problem between $r=r_{{\rm{in}}}$ and $r_{\rm{out}}$, where we impose the boundary conditions:

\begin{eqnarray*}
n_{e}(r_{{\rm{in}}})=n_{0} ,\quad
T(r_{{\rm{in}}})=T_{\rm{in}} ,
\end{eqnarray*}
\begin{eqnarray}
T(r_{{\rm{out}}})=T_{\rm{out}} ,\quad
r_{{\rm{in}}}^{2}F(r_{{\rm{in}}})=0 .\label{boundcond}\\ \nonumber
\end{eqnarray}

\noindent
The first three conditions are taken from either the best-fit models of ZN03 or the observational data (see Table \ref{table1}), while the last condition ensures that there are no sources or sinks of heat at the cluster center. Equations (\ref{conduction}), (\ref{hydro5}), (\ref{hydro6}) and boundary conditions (\ref{boundcond}) form an eigenvalue problem with either $f$, $\epsilon$ or $\dot{M}$ as the eigenvalue, provided that the other two are given as free model parameters. The results of the steady-state cluster models presented in this subsection are not sensitive to the specific choices for the value of $r_{{\rm{in}}}$ and $r_{{\rm{out}}}$; here we choose $r_{{\rm{in}}}=1$ kpc and $r_{{\rm{out}}}=1000$ kpc, unless otherwise stated. We note that the cluster stability {\it does} depend on the value of $r_{{\rm{in}}}$; this is degenerate with the choice of $T_{\rm in}$, which is explored in \S~\ref{section:depsteady}.
We have experimented with different values of $r_{{\rm{out}}}$ for our models (e.g., $r_{{\rm{out}}}$ is taken to be $300$ kpc in models of A2597; see Table \ref{table1}), and find that the results of stability analysis are not sensitive to this choice. This is consistent with the extremely long gas cooling time in the cluster outer regions. 

To illustrate our results clearly, we need to define several physical quantities for each steady-state cluster model. We first define the volume-integrated AGN heating rate as

\begin{eqnarray}
h_{{\rm{agn}}}=\int _{r_{{\rm{in}}}}^{r_{{\rm{out}}}}4\pi r^{2}\mathcal{H}dr ,\\ \nonumber
\end{eqnarray}

\noindent
and the X-ray luminosity as

\begin{eqnarray}
L_{{X}}=\int _{r_{{\rm{in}}}}^{r_{{\rm{out}}}}4\pi r^{2}\rho \mathcal{L} dr .\\ \nonumber
\end{eqnarray}

\noindent
A cluster without any heating source will lose its thermal energy by emitting X-rays. We can thus define the isobaric cooling time from equation (\ref{hydro3}) as:

\begin{eqnarray}
 t_{\rm{cool}} \equiv \frac{\gamma}{\gamma-1}\left(\frac{P}{\rho \mathcal{L}}\right) . \label{coolingtime}\\ \nonumber
\end{eqnarray}

Table \ref{table1} lists the physical parameters and results of the steady-state models for four typical cool core clusters. 
Note that the mass accretion rates in our steady-state models are usually much less than the Eddington rate $\dot{M}_{\rm{ed}}\approx 26 (M_{\rm{bh}}/10^{9}M_{\sun})(\eta/0.1)^{-1}$ $M_{\sun}$/yr, where $M_{\rm{bh}}$ is the mass of the supermassive black hole at the cluster center and $\eta$ is the radiative efficiency of AGN accretion. 
We take the cluster Abell 1795 \citep{2002MNRAS.331..635E} as our fiducial cluster. In the first two models (A1 and A2), the values of $\epsilon$ and $\dot{M}$ are given and $f$ is obtained as the eigenvalue. Model A1 is a pure conduction model ($\epsilon=0$), while model A2 is a typical hybrid model with both thermal conduction and AGN feedback heating. In model A3, the AGN feedback efficiency $\epsilon$ is chosen to be much larger than that in model A2. We further assume that, in addition to the boundary conditions (\ref{boundcond}), the electron density at the outer boundary is fixed (we choose the same as that in model A2). Thus, with the extra boundary condition, both $\dot{M}$ and $f$ can be solved as eigenvalues. The radial steady-state profiles of Abell 1795 are presented in Figure \ref{plotone}. As can clearly be seen, the electron density and temperature profiles of these three models fit the observational data quite well. The profiles of models A2 and A3 are virtually the same, since the gas density and temperature at both the inner and outer boundaries are exactly the same for these two models. Using these cluster profiles and equation (\ref{coolingtime}), we can then calculate the radial profiles of the isobaric cooling time, which are shown in the {lower left} panel of Figure \ref{plotone}. Obviously, $t_{\rm{cool}}$ is much less than the age of the Universe within $\sim 100$ kpc from the cluster center, suggesting that the radiative cooling is dynamically important in the central regions of the cluster. We show the relative importance of AGN heating and thermal conduction in the {lower right} panel of Figure \ref{plotone}. For our hybrid models, while thermal conduction is significant in the outer parts of the cluster cool core, AGN heating clearly dominates at the center. Thus the gas temperature profile in the innermost regions ($\lesssim10$ kpc) of the cluster is flatter than that in the pure conduction model, as clearly seen in the {upper right} panel of Figure \ref{plotone}. 

Since we will find later that the dependence of the cluster stability on $\epsilon$ varies somewhat with cluster properties, particularly the density and temperature profiles, we choose the cluster Abell 2199 \citep{2002MNRAS.336..299J} as our second fiducial cluster and plot its radial steady-state profiles in Figure \ref{plottwo}. The gas cooling time in its central cool core ($r \lesssim 100$ kpc) is clearly less than the Hubble time.

\begin{figure}
\epsscale{1.0}
\plotone{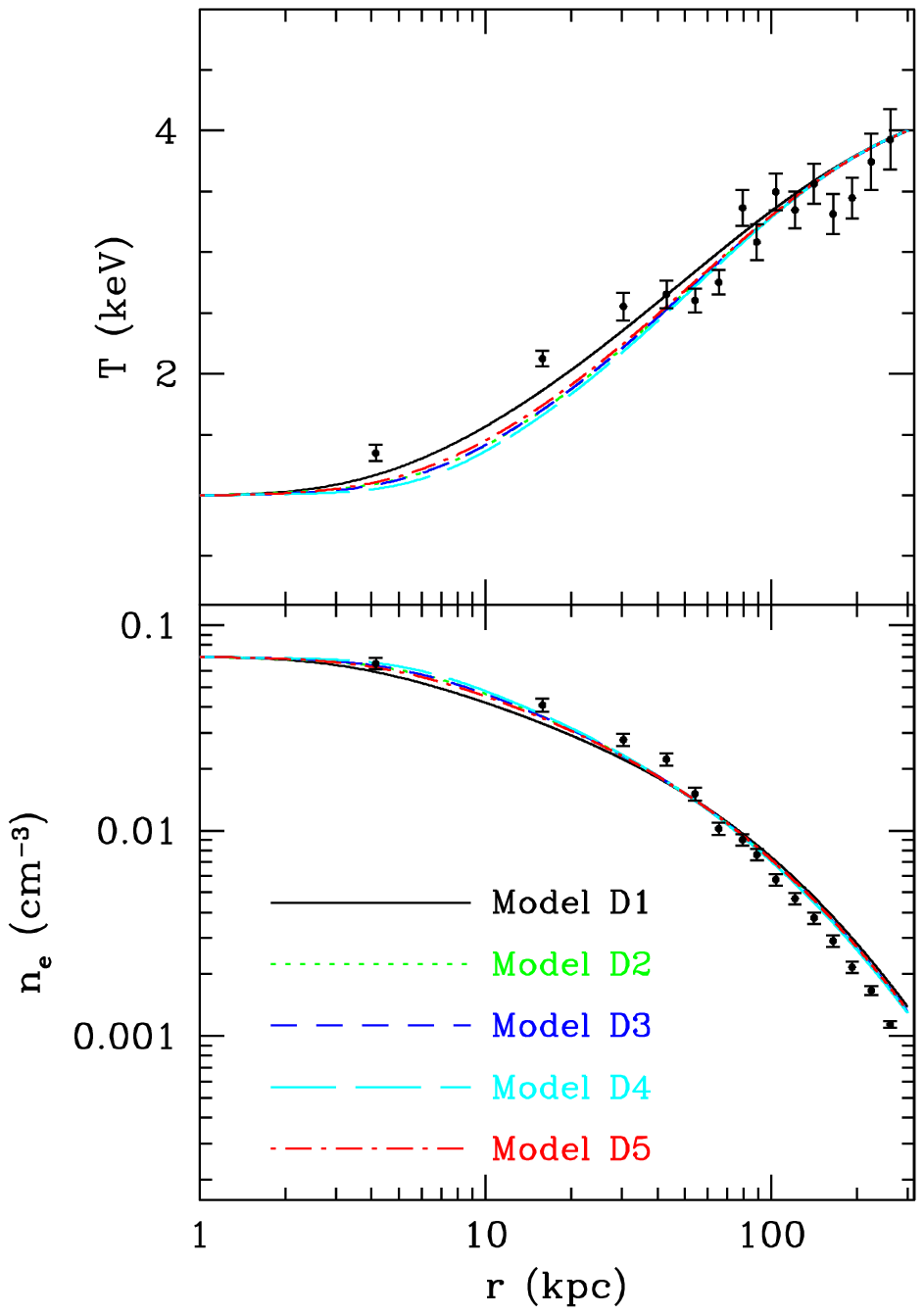}
\caption{Electron temperature (\textit{upper}) and number density (\textit{lower}) in five typical steady-state models of the cluster Abell 2597. The data points are from the \textit{Chandra} Observation \citep{2001ApJ...562L.149M}. See Table \ref{table1} for additional information.}
 \label{plot3}
 \end{figure}

ZN03 found that, for some cool core clusters, conduction-only models may require implausibly high values of thermal conductivity. We apply our model to one of these clusters, Abell 2597. As shown in Table \ref{table1}, the conduction-only model (D1) for A2597 requires a thermal conductivity of $f=1.3$. By including AGN feedback heating, the required thermal conductivity may be much smaller (e.g., $f=0.4$ in model D3). Figure \ref{plot3} shows radial profiles of electron temperature and number density in our models for A2597 (model D1 - D5), which all fit the data reasonably well. We note that some cool core clusters may not be well explained by models with conduction and AGN ``effervescent'' heating (see \citealt{2006A&A...453..423P}). However, they may be explained by models with a more realistic AGN heating function, and our method of global stability analysis is generally applicable to any steady-state AGN feedback models with only slight modifications due to the new AGN heating function.

For each cluster, the first model in Table \ref{table1} is the pure conduction model, where the cooling is entirely balanced by thermal conduction. This model determines the maximum value of $f$ for each cluster: as the level of AGN heating increases, the required thermal conductivity decreases. In the current paper, we only present our results from some typical models with specific values of $f$ (e.g., $f=0.12$ in model A3 for the cluster A1795), but the conclusions drawn are general to models with different levels of thermal conduction (also see Table 3 of \citealt{2006A&A...453..423P}). As an example, we additionally present models with various values of $f$ for the cluster A2597 (models D4 and D5) in Table \ref{table1} and Figure \ref{plot3} (also see Figure \ref{plot7}).

We note that a minimum level of thermal conduction is usually required in our steady-state models, since $\mathcal{H}/\rho \mathcal{L}$ is not uniform with radius, as readily seen in the {lower right} panels of Figure \ref{plotone} and \ref{plottwo}. In other words, the ``effervescent'' AGN heating mechanism (equation \ref{agnheat}) alone could not offset cooling at all radii throughout the cluster. Nonetheless, as we have shown, a combination of a moderate level ($f \gtrsim f_{\rm{min}}$, where $f_{\rm{min}}\sim 0.1$ for A1795 and $\sim 0.2$ for A2597) of thermal conduction and the ``effervescent'' AGN heating can balance cooling throughout the entire cluster, and the resulting steady-state profiles of electron density and temperature fit the observational data quite well. As $f$ decreases, the heating from thermal conduction decreases, and thus the required AGN heating increases. When $f\lesssim f_{\rm{min}}$, the required AGN heating in steady-state models with the boundary conditions (\ref{boundcond}) usually overheats the cluster central regions, while the gas temperature and entropy start to drop rapidly with increasing radius, which is not consistent with X-ray observations. Note that the requirement on conductivity ($f \gtrsim f_{\rm{min}}$) may be alleviated if other forms of AGN heating (e.g., viscous dissipation of sound waves, see \citealt{2004ApJ...611..158R}; shock heating, see \citealt{2007MNRAS.380L..67B}; cosmic ray feedback, see \citealt{2007arXiv0706.1274G}) are taken into account, since a substantial fraction of sound waves, shocks or cosmic rays produced at the cluster center may be dissipated in the outer regions of the ICM, with perhaps different spatial heat deposition from what we have assumed. Further work on this is clearly needed. 

\section{Radial Stability Analysis}
\label{section:stability}

\subsection{Perturbation equations}

We linearize equations (\ref{hydro1}) $-$ (\ref{conduction}) by using the Lagrangian perturbation method. The background ICM is assumed to be in steady state, as described in \S~\ref{section:stst}. A Lagrangian perturbation, denoted by an operator $\Delta$, is related to an Eulerian perturbation $\delta$ in the usual way,

\begin{eqnarray}
 \Delta=\delta+\mbox{\boldmath $\xi$} \cdot \nabla , \\ \nonumber
\end{eqnarray}

\noindent
where $\mbox{\boldmath $\xi$}$ is the displacement vector of a fluid element \citep[see, e.g.,][pp. 130-147]{1983bhwd.book.....S}. By perturbing equations (\ref{hydro1}) and (\ref{estate}), we find

\begin{eqnarray}
 \Delta \rho=-\rho \nabla \cdot  \mbox{\boldmath $\xi$} , \label{pert1}
\end{eqnarray}
\begin{eqnarray}
 \Delta P=P\frac{\Delta T}{T}-P \nabla \cdot \mbox{\boldmath $\xi$} . \\ \nonumber
\end{eqnarray}

\noindent
Here we only consider radial perturbations and 

\begin{eqnarray}
 \nabla \cdot \mbox{\boldmath $\xi$}=\frac{1}{r^{2}}\frac{\partial}{\partial r}(r^{2}\xi) , \label{pert1a}\\ \nonumber
\end{eqnarray}

\noindent
where $\xi=\Delta r$ denotes the radial component of $\mbox{\boldmath $\xi$}$. To derive the perturbation equations, we use the following properties of $\Delta$ \citep{1983bhwd.book.....S}:

\begin{eqnarray}
 \Delta \frac{d}{dt}=\frac{d}{dt}\Delta ,  
\end{eqnarray}
\begin{eqnarray}
 \Delta \frac{\partial}{\partial r}=\frac{\partial}{\partial r}\Delta-\frac{\partial \xi}{\partial r}\frac{\partial}{\partial r}
 .  \label{pert2}\\ \nonumber
\end{eqnarray}

\noindent
Applying $\Delta$ to equations (\ref{hydro2}) $-$ (\ref{conduction}) and using equations (\ref{hydro3}) and (\ref{pert1}) - (\ref{pert2}), we obtain

\begin{eqnarray}
 \frac{d^{2}\xi}{d t^{2}}=\frac{P}{\rho}\frac{\partial}{\partial r}(\nabla \cdot \mbox{\boldmath $\xi$})-\frac{1}{\rho} 
 \frac{\partial}{\partial r}\left(P\frac{\Delta T}{T}\right)+\frac{1}{\rho}  \frac{dP}{dr}\frac{\partial \xi}{\partial r}
 -\xi \frac{d^{2} \Phi}{dr^{2}}
 ,\label{stab1}
\end{eqnarray}

\begin{eqnarray}
\kappa T \frac{\partial}{\partial r}\left(\frac{\Delta T}{T}\right) = F\left(\frac{7}{2}\frac{\Delta T}{T}-\frac{\partial \xi}{\partial r}+ \frac{2\xi}{r}\right)+\frac{\Delta L_{r}}{4\pi r^{2}}
  ,\label{stab2}
\end{eqnarray}

\begin{eqnarray}
\frac{1}{4\pi r^{2}}\frac{\partial}{\partial r}\Delta L_{r} =  \left(P\frac{d}{dt}-\rho^{2}\mathcal{L}_{\rho}-\mathcal{H}\right)(\nabla \cdot \mbox{\boldmath $\xi$})-\Delta \mathcal{H} & & \nonumber \\ 
+ \left(\frac{P}{\gamma-1}\frac{d}{dt}+\rho T \mathcal{L}_{T}+\frac{1}{\gamma-1}\frac{dP}{dt}-\frac{\gamma}{\gamma-1}\frac{P}{\rho}\frac{d\rho}{dt}\right)\frac{\Delta T}{T}, & &\label{stab3}\\ \nonumber
\end{eqnarray}

\noindent
where $\mathcal{L}_{T} \equiv \partial \mathcal{L}/\partial T|_{\rho}$, $\mathcal{L}_{\rho} \equiv \partial \mathcal{L}/\partial \rho|_{T}$, $L_{r}=-4\pi r^{2}F$ is the radial heat luminosity, and $\Delta \mathcal{H}$ is the perturbation of AGN heating rate (equation \ref{agnheat})

\begin{eqnarray}
\Delta \mathcal{H} & = &\left(-2+\frac{r}{r_{0}}\frac{e^{-r/r_{0}}}{1-e^{-r/r_{0}}}\right)\frac{\xi}{r}\mathcal{H} +\frac{\mathcal{H}}{\partial P/\partial r}\frac{\partial}{\partial r} \Delta P -\mathcal{H}\frac{\partial \xi}{\partial r} \nonumber \\
& + & \mathcal{H}\left[(\beta-1)\frac{\Delta P}{P}-\beta \frac{\Delta P_{0}}{P_{0}}\right] + \Delta \mathcal{H}_{\rm{feed}},\label{stab4}\\ \nonumber 
\end{eqnarray}

\noindent
where $\Delta P_{0}$ is the value of $\Delta P$ at the inner boundary $r_{\rm{in}}$, and $\Delta \mathcal{H}_{\rm{feed}} \equiv  \mathcal{H}\Delta \dot{M}(r_{\rm{in}})/\dot{M}_{\rm{in}}$
is the perturbation of the AGN heating rate due to the feedback mechanism, where $\Delta \dot{M}(r_{\rm{in}})$ is the perturbation of the mass accretion rate at $r=r_{\rm{in}}$,

\begin{eqnarray}
 \Delta \dot{M}(r_{\rm{in}})=\frac{\dot{M}_{\rm{in}}}{v_{0}}\frac{\partial \xi}{\partial t}(r_{\rm{in}}) ,
\label{pertfeed}\\ \nonumber
\end{eqnarray}

\noindent
which can be easily derived from perturbing the definition of the mass accretion rate ($\dot{M}\equiv 4\pi r^{2}\rho v$) and using equations (\ref{pert1}) and (\ref{pert1a}).
Note that we have neglected any time delay between central AGN activity and the resulting heating of the ICM, which is justifiable since both the AGN duty cycle and the bubble rising time are usually much shorter than the gas cooling time (see the discussion above equation \ref{agnheat}).

Taking $\xi$, $\Delta T$, $\Delta L_{r}$ as independent variables, we seek solutions of equations (\ref{stab1}) $-$ (\ref{stab3}) that behave as $\sim e^{\sigma t}$ with time. The term $\Delta \mathcal{H}_{\rm{feed}}$ in equation (\ref{stab4}) then simplifies to

\begin{eqnarray}
\Delta \mathcal{H}_{\rm{feed}} =\frac{\mathcal{H} \sigma}{v_{0}} \xi (r_{\rm{in}}) ,
\label{delta_H_feed_simple} \\ \nonumber
\end{eqnarray}

\noindent
and equations (\ref{stab1}) $-$ (\ref{stab3}) may be rewritten as

\begin{eqnarray}
\left(\frac{P}{\rho}-v^{2}\right)\frac{d}{dr}(\nabla \cdot \mbox{\boldmath $\xi$})=
 \left(r\sigma^{2}+r\frac{d^{2} \Phi}{dr^{2}}\right)\frac{\xi}{r}+\frac{1}{\rho} 
 \frac{d}{dr}\left(P\frac{\Delta T}{T}\right) & & \nonumber\\
 -2v^{2}\frac{d}{dr}\left(\frac{\xi}{r}\right)+ \left(2\sigma v+v\frac{dv}{dr}-\frac{1}{\rho}\frac{dP}{dr}\right) \frac{d\xi}{dr}, & & \label{gstab1}
\end{eqnarray}

\begin{eqnarray}
\kappa T \frac{d}{dr}\left(\frac{\Delta T}{T}\right) = F\left[\frac{7}{2}\frac{\Delta T}{T}-r\frac{d}{dr}\left(\frac{\xi}{r}\right)+ \frac{\xi}{r}\right]+\frac{\Delta L_{r}}{4\pi r^{2}}
  ,\label{gstab2}
\end{eqnarray}

\begin{eqnarray}
 \frac{1}{4\pi r^{2}}\frac{d}{dr}\Delta L_{r} = (P\sigma-\rho^{2}\mathcal{L}_{\rho}-\mathcal{H})(\nabla \cdot \mbox{\boldmath $\xi$})-\Delta \mathcal{H}  & & \nonumber \\ 
 + \left(\frac{P\sigma}{\gamma-1}  +\rho T \mathcal{L}_{T}+\frac{v}{\gamma-1}\frac{dP}{dr}-\frac{\gamma v}{\gamma-1}\frac{P}{\rho}\frac{d\rho}{dr}\right)\frac{\Delta T}{T} & &  \nonumber \\
 + Pv\frac{d}{dr}(\nabla \cdot \mbox{\boldmath $\xi$})+ \frac{Pv}{\gamma-1}\frac{d}{dr}\left(\frac{\Delta T}{T}\right)
   .\label{gstab3}\\ \nonumber
\end{eqnarray}

\noindent
In equations (\ref{gstab1}) $-$ (\ref{gstab3}) and hereinafter, we omit $e^{\sigma t}$ from all perturbation variables. 

Equations (\ref{gstab1}) $-$ (\ref{gstab3}) form an eigenvalue problem, which can be solved numerically with appropriate boundary conditions to find global eigenmodes and the eigenvalue $\sigma$. Before considering the global modes, in the following subsection, we first study small-scale local modes, which may not be captured in a global stability analysis (see the discussion in \citealt{1989ApJ...341..611B}). The nature of local thermal stability in a stratified system can be subtle, and linked to the convective instability of the system (Balbus 1988).

\subsection{Local stability analysis of radial modes}
\label{localrad}
 
For local stability analysis, we consider local WKB perturbations of the form $\sim e^{ik_{r}r+\sigma t}$. 
Here we neglect the feedback mechanism of AGN heating, i.e. $\Delta \mathcal{H}_{\rm{feed}}$ in equation (\ref{stab4}) is taken to be zero. This term simply affects the overall normalization of heating without spatial dependence, and is only important in a global analysis. The subsonic background flow is ignored as well, so that the unperturbed steady-state ICM is nearly in hydrostatic equilibrium. In the local approximation, we assume that $k_{r}r \gg 1$ (plane-parallel approximation) and that the wavelengths of perturbations are much shorter than any spatial scale on which the background quantities vary (e.g., $k_{r}\gg 1/\lambda_{p}$, where $\lambda_{p}\equiv (d \ln P/dr)^{-1}$ is the gas pressure scale height in the ICM). To eliminate high-frequency sound waves from consideration, we also assume that $|\sigma| \ll c_{s}k_{r}$, where $c_{s}=\sqrt{P/\rho}$ is the isothermal sound speed. Therefore, the perturbed dynamical equation of motion (equation \ref{gstab1}) simplifies to (KN03):

\begin{eqnarray}
ik_{r}\xi=\frac{\Delta T}{T}
  .  \label{kdeltat}\\ \nonumber
\end{eqnarray}
(note that in general, $\Delta T/T$ is complex). 

Similarly, the perturbed energy equation (\ref{gstab3}) may be rewritten as

\begin{eqnarray}
(P\sigma-\rho^{2}\mathcal{L}_{\rho}-\rho \mathcal{L})ik_{r}\xi+\left(\frac{P\sigma}{\gamma-1}  +\rho T \mathcal{L}_{T}\right)\frac{\Delta T}{T} \nonumber & &\\
 = \Delta \left(\frac{1}{4\pi r^{2}}\frac{d}{dr}L_{r}\right)+\Delta \mathcal{H}. & &\label{perturbe}\\ \nonumber
\end{eqnarray}

\noindent 
Taking the leading order terms from the perturbations of thermal conduction and AGN heating, we obtain

\begin{eqnarray}
 \Delta \left(\frac{1}{4\pi r^{2}}\frac{d}{dr}L_{r}\right)=-\kappa T k_{r}^{2}\frac{\Delta T}{T} ,  
\end{eqnarray} 

\begin{eqnarray}
 \Delta \mathcal{H} =\frac{\mathcal{H}}{\partial P/\partial r}\Delta \frac{\partial P}{\partial r}  =-\mathcal{H}ik_{r}\xi
  .  \label{deltahl}\\ \nonumber
\end{eqnarray}

\noindent 
In the second equality of equation (\ref{deltahl}), the perturbation of the radial pressure gradient, $\Delta (\partial P/\partial r)$, is evaluated by perturbing the Euler equation (equation (\ref{hydro2})) and assuming that $\sigma^{2}\ll k_{r}c_{s}^{2}/\lambda_{p}$, which corresponds to slowly evolving perturbations\footnote{Provided that $|\sigma| \ll c_{s}k_{r}$, $\sigma^{2}\ll k_{r}c_{s}^{2}/\lambda_{p}$ is guaranteed as long as $|\sigma|<c_{s}/\lambda_{p}$ (i.e., the growth time of the perturbation is longer than the sound crossing time over a pressure scale height).}.

Equations (\ref{kdeltat}) $-$ (\ref{deltahl}) may be combined to give

\begin{eqnarray}
\sigma=\sigma_{\infty}-\frac{\gamma-1}{\gamma}\frac{\mathcal{H}}{P}
-\frac{\gamma-1}{\gamma}\frac{\kappa T}{P}k_{r}^{2}
 ,   \label{localrate}\\ \nonumber
\end{eqnarray} 

\noindent
where 

\begin{eqnarray}
\sigma_{\infty}=-\frac{\gamma-1}{\gamma}\frac{\rho T^{2}}{P}\left(\frac{\partial \mathcal{L}/T}{\partial T}\right)_{P} \\ \nonumber 
\end{eqnarray}

\noindent 
is the growth rate of local isobaric thermal instability in the ICM without any heating (e.g., \citealt{1965ApJ...142..531F}; KN03). In the absence of any heating source, equation (\ref{localrate}) reduces to $\sigma=\sigma_{\infty}$, and we thus immediately recover the generalized Field criterion for isobaric thermal instability \citep{1986ApJ...303L..79B}

\begin{eqnarray}
\left(\frac{\partial \mathcal{L}/T}{\partial T}\right)_{P} <0 . \label{localcrit}\\ \nonumber
\end{eqnarray} 

\noindent
For the ICM with $\mathcal{L}\propto \rho T^{1/2}$, the instability criterion given by equation (\ref{localcrit}) is easily met, suggesting that local radial perturbations grow exponentially with the growth time

\begin{eqnarray}
t_{\infty} & \equiv & \sigma_{\infty}^{-1}\nonumber \\
           & =      & 0.64\; {\rm Gyr}\left(\frac{n_{e}}{0.05\rm{ cm}^{-3}}\right)^{-1}\left(\frac{k_{B}T}{2\rm{ keV}}\right)^{1/2}.\label{tinfty}\\ \nonumber
 \end{eqnarray}

Equation (\ref{localrate}) confirms the well known result that thermal conduction stabilizes short-wavelength perturbations (e.g., \citealt{1965ApJ...142..531F}; \citealt{1987ApJ...319..632M}; KN03). More specifically, for the ICM with $\mathcal{L}\propto \rho T^{1/2}$, thermal conduction alone stabilizes perturbations whose wavelengths ($\lambda \equiv 2\pi/k_{r}$) are smaller than the critical wavelength

\begin{eqnarray}
\lambda_{\rm{Field}} & = & 2\pi \left(\frac{2\kappa T}{3\rho \mathcal{L}}\right)^{1/2} \nonumber\\
                       & = & 25.6\; {\rm kpc}\left(\frac{f}{0.2}\right)^{1/2}\left(\frac{n_{e}}{0.05\rm{ cm}^{-3}}\right)^{-1}\left(\frac{k_{B}T}{2\rm{ keV}}\right)^{3/2}. \label{lambdacr}\\ \nonumber
\end{eqnarray} 

Equation (\ref{localrate}) also clearly shows that an AGN heating term as implemented in our model ($\mathcal{H} \propto \partial P/\partial r$) reduces the growth rate of local thermal instability, although it alone could not suppress local thermal instability completely 
(note that $(\gamma-1)\mathcal{H}/(\gamma P\sigma_{\infty})=2\mathcal{H}/3\rho\mathcal{L} \le 2/3$, since $\mathcal{H} \le \rho \mathcal{L}$). We note that local thermal instability of the ICM may depend on the actual mechanism by means of which the AGN mechanical energy is transferred to the thermal ICM; we have assumed complete local dissipation of the $pdV$ work done by the expanding bubbles. Alternative dissipation mechanisms (e.g., sounds waves which damp far away, or the heating effect of dispersed cosmic rays) could yield different results. Furthermore, in our 1D model we have assumed isotropic heating by bubbles; in reality the angular variation of bubble creation will affect local (and global) thermal stability. 

The local stability analysis is not valid for long-wavelength perturbations, and a successful model for the ICM must be globally stable. KN03 studied the global stability of the pure conduction model, and found that it is globally unstable with the typical instability growth time of $\sim 2-5$ Gyr; the growth time can be significantly shorter if one applies non-linear perturbations. In the next subsection, we will perform a global stability analysis for our steady-state cluster models presented in \S~\ref{section:stst}, and show that the feedback mechanism is essential to stabilize global thermal instability in cool core clusters. 
 
\subsection{Global unstable modes}
\label{section:globalstability}

\begin{table}
 \centering
 \begin{minipage}{80mm}
  \caption{Timescales for the clusters shown in Table~\ref{table1}}
   \vspace{3mm} 
  \begin{tabular}{@{}lccccc}
  \hline
  \hline
       & {$t_{\rm{cool},0}$\footnote{$t_{\rm{cool},0}$ is the isobaric cooling time at the cluster center.}} & 
        {$t_{\infty,0}$ \footnote{$t_{\infty,0}$ is the growth time of the local isobaric thermal instability at the cluster center in absence of any heating source (see equation \ref{tinfty}).}} & && {$t_{\rm{grow}}$ \footnote{$t_{\rm{grow}}$ is the growth time of the unstable global radial mode.}}\\
     Name& (Gyr)& (Gyr)&{$\epsilon_{\rm{min}}$\footnote{$\epsilon_{\rm{min}}$ is the lower limit of the AGN feedback efficiency, above which the ICM is effectively ($t_{\rm{grow}}>t_{H}$) or completely stable.}}&Model&(Gyr)       \\
 \hline
  A1795  & 0.9&0.6& 0.28&A1  &3.8  \\
 &   & &&A2& {3.3, 43.3 \footnote{There are two unstable global radial modes in this model.}}   \\
  &&& &A3 &stable  \\
 A2199 &  0.6 & 0.4  &0.17&B1 & 2.8\\
 &&& &B2 &  4.4\\
  &&& &B3 &  16.9\\
 A2052 & 1.1  & 0.7&0.14&C1 & 6.2  \\
    &&&& C2&5.9  \\
      &&& &C3 & 20.0 \\  
 A2597 & 0.5  & 0.3 &0.07& D1&2.0 \\
     &&&& D2& 3.3 \\
      &&& &D3& 38.1  \\         
     &&&& D4& 27.5 \\
      &&& &D5& 20.3  \\  
\hline
\label{table2}
\end{tabular}
\end{minipage}
\end{table}

\begin{figure}
\epsscale{1.0}
\plotone{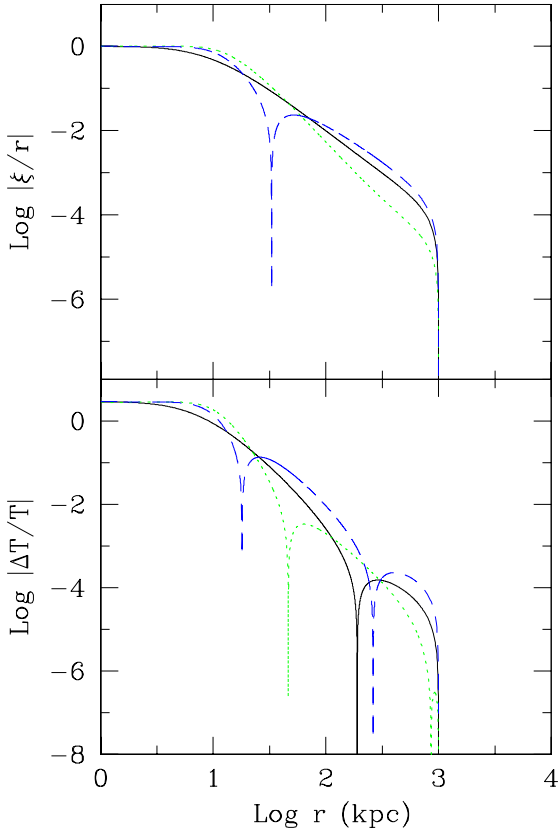}
\caption{Eigenfunctions of the radial unstable modes for the steady-state models of Abell 1795 presented in Fig. \ref{plotone}, plotted as a function of radius. The solid lines stand for the lone unstable mode for the pure conduction model (model A1 in Table \ref{table1}). Model A2 has two unstable modes: $t_{\rm{grow}}=3.3$ Gyr (dotted lines) and $t_{\rm{grow}}=43.3$ Gyr (dashed lines). Model A3 has no unstable radial modes.}
 \label{plot4}
 \end{figure}
 
 \begin{figure}
 \epsscale{1.0}
\plotone{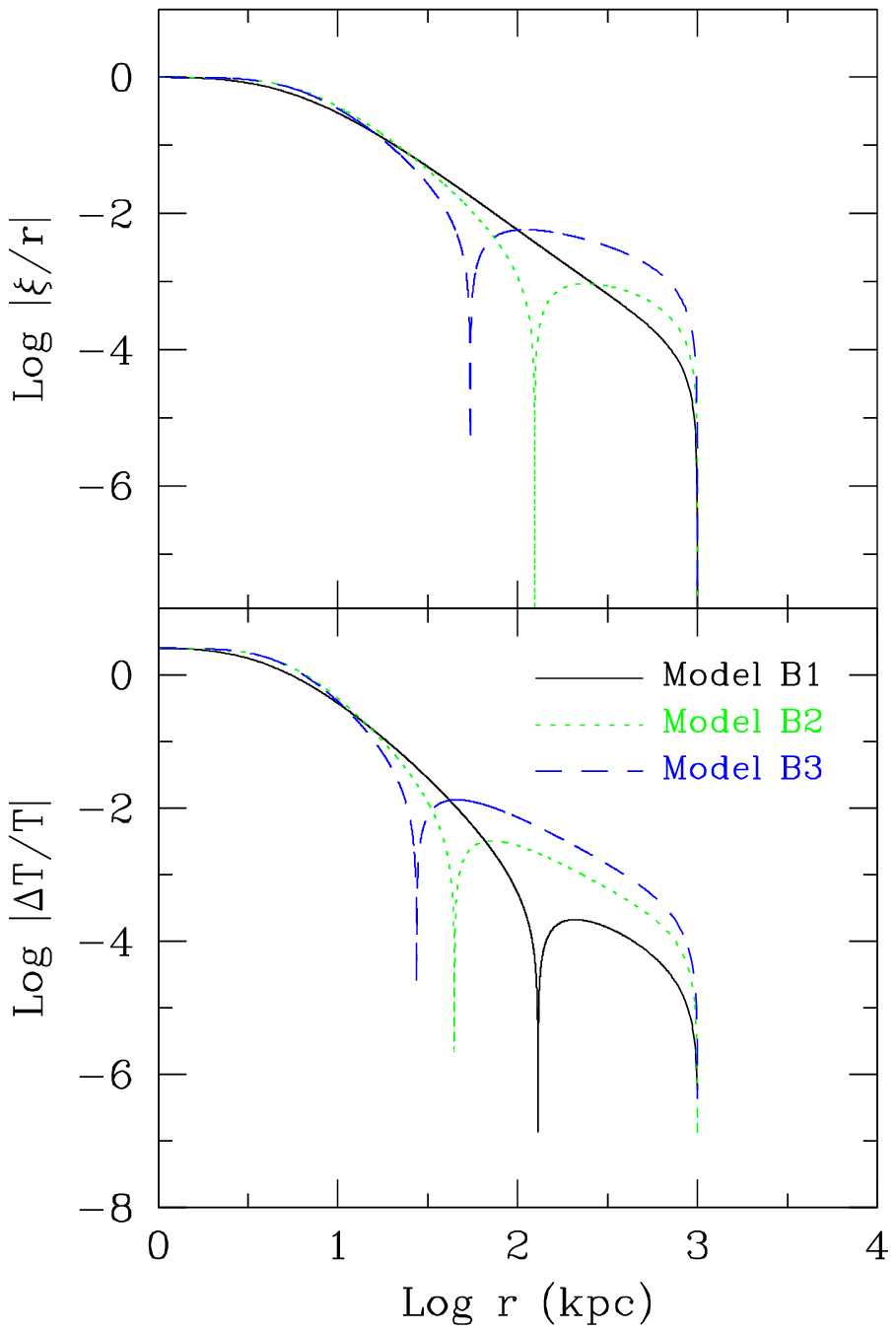}
\caption{Eigenfunctions of the radial unstable modes for the steady-state models of Abell 2199 presented in Fig. \ref{plottwo}, plotted as a function of radius. Each model has one unstable radial mode.}
 \label{plot5}
 \end{figure}

To analyze global stability of the steady-state models of a given cluster, we numerically solve
equations (\ref{gstab1}) $-$ (\ref{gstab3}), which are equivalent to four first-order differential equations for the four variables $\xi/r$, $\Delta T/T$, $\Delta L_{r}$ and $\nabla \cdot \mbox{\boldmath $\xi$} =3\xi/r+rd(\xi/r)/dr$. We solve these equations as an eigenvalue problem, where the eigenvalue is the growth rate $\sigma$. Since we have four variables and one eigenvalue, we need to specify five boundary conditions. Here we choose the same boundary conditions as KN03. The three inner boundary conditions are 

\begin{eqnarray}
 \frac{\xi}{r}=1
 ,\quad  \frac{d}{dr}\left( \frac{\xi}{r}\right)=0
 ,\quad  \Delta L_{r}=0 ,\quad {\rm at}\; r=r_{\rm in}  .\\ \nonumber
\end{eqnarray}

\noindent
The first condition is a normalization condition, while the second condition guarantees that the solutions are regular (due to the presence of a $(1/r)d/dr(\xi/r)$ term in the simplified form of equation \ref{gstab1}).
Since $r_{\rm in}$ is not exactly zero, the second condition need not hold strictly. We have experimented with different values for the second condition (for instance, $d/dr(\xi/r) = \xi/r^{2}$), and find that the results are not sensitive to it. This is consistent with the fact that the contribution of AGN feedback (equation \ref{delta_H_feed_simple}) in the perturbed energy equation (\ref{gstab3}) is independent of this condition, although the exact value of the instability growth time for each model will be slightly affected due to its contribution to the perturbed cooling term ($P\sigma \nabla \cdot \mbox{\boldmath $\xi$}$) in equation (\ref{gstab3}).
The last condition demands that the perturbed radial heat luminosity is zero at the cluster center, which is equivalent to a zero temperature gradient there. The remaining two outer boundary conditions are set by the requirement that perturbations vanish at the outer boundary, which has cooling times much longer than the cluster lifetime:

\begin{eqnarray}
\xi=0
 , \quad \Delta T=0
 , \quad {\rm at}\; r=r_{\rm out } . \label{outbound} \\ \nonumber
\end{eqnarray}
 
We use the steady-state models constructed in \S~\ref{section:stst} as the background states to calculate the eigenmodes of global perturbations. Similar to KN03, we first fix $\sigma$ and set $\Delta T/T$ to an arbitrary value at $r=r_{{\rm{in}}}$. We can then integrate eqautions (\ref{gstab1}) $-$ (\ref{gstab3}) from $r=r_{{\rm{in}}}$ to $r=r_{\rm{out}}$ using a Runge-Kutta method. We use the bisection method to update the inner value of $\Delta T/T$ and continue iterating until the first outer boundary condition in equation (\ref{outbound}) is satisfied. Finally, we scan $\sigma$ in the range $(10^{4}\rm{ Gyr})^{-1}<\sigma<(10^{-4}\rm{ Gyr})^{-1}$, and use the second condition in equation (\ref{outbound}) as a discriminant for solutions. 

We first study eigenmodes with a real positive $\sigma$, which correspond to globally unstable modes with an instability growth time $t_{\rm{grow}}=1/\sigma$. The results for the steady-state models listed in Table \ref{table1} are shown in Table 2, where, for comparison, we also list the central values of the isobaric cooling time (equation \ref{coolingtime}) and the growth time $t_{\infty}$ (equation \ref{tinfty}) of local isobaric thermal instability in absence of any heating source. For our fiducial cluster Abell 1795, the pure conduction model A1 ($f=0.27$) has one unstable mode with growth time $t_{\rm{grow}}=3.8$ Gyr, which is consistent with KN03, who found that the equilibrium model of A1795 ($f=0.2$) has one unstable mode with $t_{\rm{grow}}=4.1$ Gyr. For model A2, where the AGN feedback efficiency is $\epsilon=0.1$, we found two unstable modes with $t_{\rm{grow}}=3.3$ Gyr and $43.3$ Gyr respectively. Model A3 has virtually the same background ICM profiles as model A2 (see \S~\ref{section:stst}), but a higher AGN feedback efficiency ($\epsilon=0.3$). Our calculation shows that model A3 has no unstable modes. This is a remarkable result, since we, for the first time, show from a linear analysis that AGN feedback completely eliminates (radial) global thermal instability. We note that, without introducing the feedback mechanism for the AGN heating (i.e., if $\Delta \mathcal{H}_{\rm{feed}}$ in equation \ref{stab4} is taken to be zero), model A3 is still unstable and has two unstable modes with $t_{\rm{grow}}=2.7$ Gyr and $52.8$ Gyr respectively. We repeated our calculations for many different values of model parameters, and found that the models without the feedback mechanism are always globally unstable, even when the same models with the feedback mechanism included are globally stable. Thus, the feedback mechanism of AGN heating is the key ingredient for stabilizing global thermal instability in cool core clusters. 

For the cluster Abell 2199, the pure conduction model (B1) is globally unstable with $t_{\rm{grow}}=2.8$ Gyr, which is consistent with KN03. With both conduction and AGN feedback heating included, models B2 and B3 are still globally unstable. However, the growth time of the unstable mode in model B3 ($\epsilon=0.2$) is $t_{\rm{grow}}=16.9 \rm{ Gyr} >t_{H}$, which suggests that thermal instability is dynamically unimportant and thus is ``effectively" suppressed. Note that, without introducing the feedback mechanism for the AGN heating, the growth time of the unstable mode in model B3 is much shorter ($t_{\rm{grow}}=2.2 \rm{ Gyr}$).

In Figure \ref{plot4} and \ref{plot5}, we plot the eigenfunctions of the unstable modes in galaxy clusters A1795 and A2199, respectively. Clearly, the perturbations have largest amplitude in the cluster central regions ($r\lesssim 10$ kpc) and decay rapidly with increasing radius, which suggests that perturbations reach nonlinear amplitudes much faster in central regions than in outer regions. Such global modes have also been found by KN03 for equilibrium ICM models with thermal conduction only. 

As shown in Table 2, we obtain similar results for the other two cool core clusters listed in Table \ref{table1}. It is worth mentioning that the pure conduction model for the cluster Abell 2597 requires an implausibly high value of the thermal conductivity ($f=1.3$) and is globally unstable. By including AGN feedback heating, model D3 requires a smaller conductivity ($f=0.4$) and is effectively stable ($t_{\rm{grow}}=38.1 \rm{ Gyr}$, for $\epsilon=0.1$).

\subsection{Dependence on the AGN feedback efficiency}
\label{section:deppara}

\begin{figure}
\epsscale{1.0}
\plotone{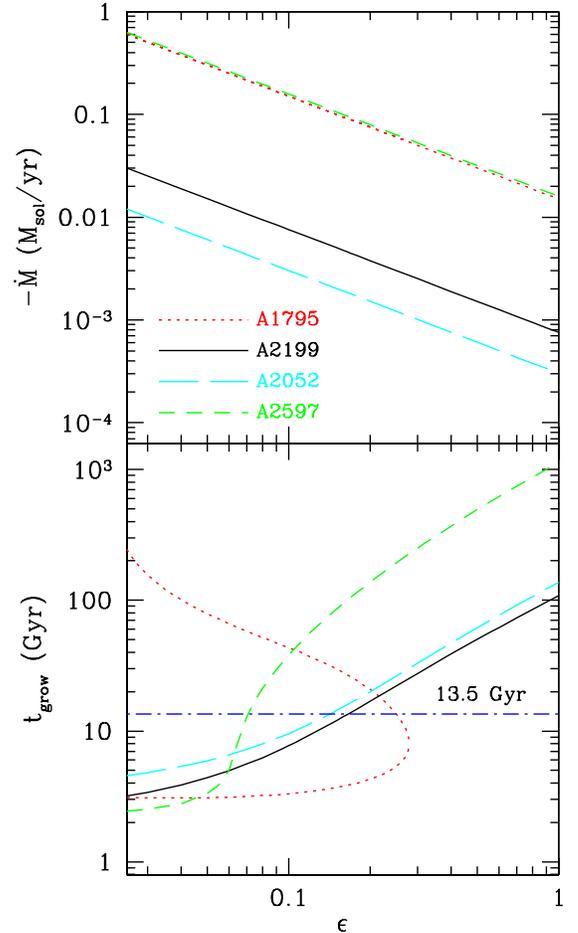}
\caption{Effect of the AGN feedback efficiency on thermal stability in typical cool core clusters. For different models of each cluster, the values of $f$ and $\epsilon \dot{M}$ are roughly the same (equal to those in the second model listed in Table \ref{table1}; see the text for details). Top panel: scaling of $\dot{M}$ with $\epsilon$. Bottom panel: the dependence of the growth time of unstable modes on $\epsilon$ for each cluster. Note that the clusters are stable or effectively stable when $\epsilon$ is greater than a lower limit $\epsilon_{\rm min}$. The line for A1795 is double-valued since it has two unstable modes; both these modes vanish when $\epsilon > 0.28$. Here, we assume $\epsilon$ is a constant; if, as suggested by observations, $\epsilon \propto \dot{M}^{\nu}$ (with $\nu \sim 0.3-0.6$), then $\epsilon_{\rm min}$ will be smaller (see text).}
 \label{plot6}
 \end{figure}

\begin{figure}
\epsscale{1.0}
\plotone{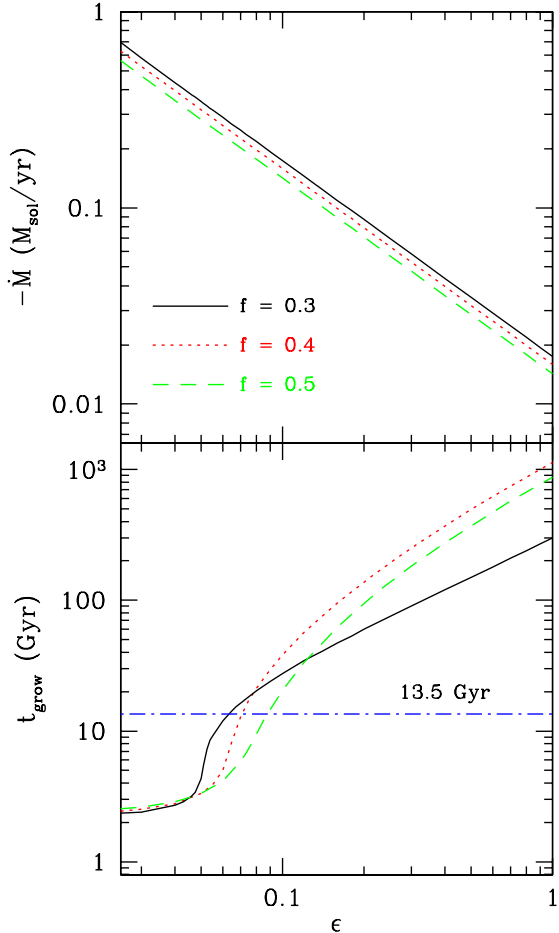}
\caption{Effect of the AGN feedback efficiency on thermal stability in A2597. Each curve is plotted as a function of $\epsilon$ with a fixed conductivity suppression factor $f$. Top panel: scaling of $\dot{M}$ with $\epsilon$. Bottom panel: the dependence of the growth time of unstable modes on $\epsilon$. For different levels of fixed conductivity, the cluster always becomes more stable as the AGN feedback efficiency increases.}
 \label{plot7}
 \end{figure}
 
From the results of the previous subsection, one might conjecture that the ICM is stable (or effectively stable) if the AGN feedback efficiency $\epsilon$ is greater than a lower limit. To fully explore the dependence of the cluster stability on the AGN feedback efficiency, we consider the steady-state cluster models subject to the following boundary conditions

\begin{eqnarray*}
n_{e}(r_{{\rm{in}}})=n_{0},\quad
T(r_{{\rm{in}}})=T_{\rm{in}},\quad
r_{{\rm{in}}}^{2}F(r_{{\rm{in}}})=0,
\end{eqnarray*}
\begin{eqnarray}
T(r_{{\rm{out}}})=T_{\rm{out}},\quad
n_{e}(r_{{\rm{out}}})=n_{\rm{out}},\label{boundcond2}\\ \nonumber
\end{eqnarray}

\noindent
where the value of $n_{\rm{out}}$ for each cluster is chosen to be that of the second model for that cluster, as listed in Table \ref{table1}. As described in \S~\ref{section:stst}, such boundary conditions ensure that the steady-state ICM profiles are almost exactly the same for models with different values of $\epsilon$ (see the lines for models A2 and A3 in Figure \ref{plotone}); it is only meaningful to study the dependence of the cluster stability on $\epsilon$ when the background profiles are virtually the same. Since the steady state cluster model only has three variables $P(r), T(r), r^{2}F(r)$ (\S~\ref{section:stst}), the five boundary conditions (\ref{boundcond2}) can determine two eigenvalues. Although our hybrid cluster models formally have three parameters ($f$, $\epsilon$ and $\dot{M}$), they are essentially determined by two parameters: $f$, which determines the level of thermal conduction, and $\epsilon \dot{M}$, which determines the level of AGN heating, while the subsonic inflow itself ($\dot{M}$) only has a negligible effect. Thus, for each cluster, the values of $f$ and $\epsilon \dot{M}$ are roughly fixed by the boundary conditions (\ref{boundcond2}), and $\dot{M}$ varies with the free parameter $\epsilon$ (see Figure \ref{plot6}).

For each steady-state cluster model, we then repeat our stability calculations and search for unstable modes. The results are shown in Figure \ref{plot6}, where the steady-state mass accretion rate and the growth time of unstable modes are plotted as a function of the AGN feedback efficiency. The value of $f$ for each cluster roughly equals to that in the second model of that cluster listed in Table \ref{table1}. The lower panel of Figure \ref{plot6} clearly shows that the cluster is stable or effectively stable ($t_{\rm{grow}} >t_{H}$) when $\epsilon$ is greater than a lower limit $\epsilon_{\rm{min}}$. We note that this result generalizes as well to models with different values of $f$, as seen in Figure \ref{plot7}, which shows the effect of AGN feedback efficiency on global stability for models of A2597 with $f=0.3, 0.4$ and $0.5$. 
Assuming that the real intracluster gas is in a stable quasi-steady state, our global stability analysis thus suggests a constraint on the kinetic efficiency of AGN feedback. As listed in Table 2, the values of $\epsilon_{\rm{min}}$ for these four typical cool core clusters are $\epsilon_{\rm{min}}\sim 0.07-0.28$, which is roughly consistent with the recent estimate of $\epsilon \sim 0.3$ for radio-loud AGNs by \citet{2007ApJ...658L...9H} and is marginally consistent with observational estimates of $\epsilon \sim 0.01-0.1$ by \citet{2006MNRAS.372...21A} and \citet{2007MNRAS.381..589M}. 
In our AGN feedback model, we assume that $\epsilon$ is a constant. However, recent X-ray observations seem to suggest that $\epsilon \propto \dot{M}^{\nu}$, where $\nu \sim 0.3$ \citep{2006MNRAS.372...21A} or $0.6$ \citep{2007MNRAS.381..589M}. In this case, the AGN feedback will be even stronger (i.e., in equation \ref{stab4}, $\Delta \mathcal{H}_{\rm{feed}}= \mathcal{H}\Delta \dot{M}(r_{\rm{in}})/\dot{M}_{\rm{in}}+\mathcal{H}\Delta \epsilon/\epsilon$), and thus
the value of $\epsilon_{\rm{min}}$ (evaluated at $\dot{M}$ appropriate for the steady-state case) will be reduced. For the cluster A1795, $\epsilon_{\rm{min}}$ is reduced from $0.28$ to $0.22$ ($\nu = 0.3$) or $0.18$ ($\nu = 0.6$). For the cluster A2199, $\epsilon_{\rm{min}}$ is reduced from $0.17$ to $0.13$ ($\nu = 0.3$) or $0.10$ ($\nu = 0.6$). 
The actual value of $\epsilon_{\rm min}$ is also sensitive to the exact form of the AGN heating law adopted, particularly its spatial dependence. Here we have only considered the $p{\rm d}V$ work due by rising bubbles, and ignored, for instance, cosmic ray heating \citep {2007arXiv0706.1274G}, viscous dissipation of sound waves \citep {2004ApJ...611..158R} or shock heating \citep{2007MNRAS.380L..67B}. However, while we do not place great store in the absolute value we obtain for $\epsilon_{\rm min}$, the fact that there is a minimal heating efficiency $\epsilon_{\rm min}$ for a given temperature profile should be fairly robust, since it only requires that $L_{\rm agn} \propto \epsilon \dot{M}$. 

\subsection{Dependence on the background profiles}
\label{section:depsteady}
 
\begin{figure}
\epsscale{1.0}
\plotone{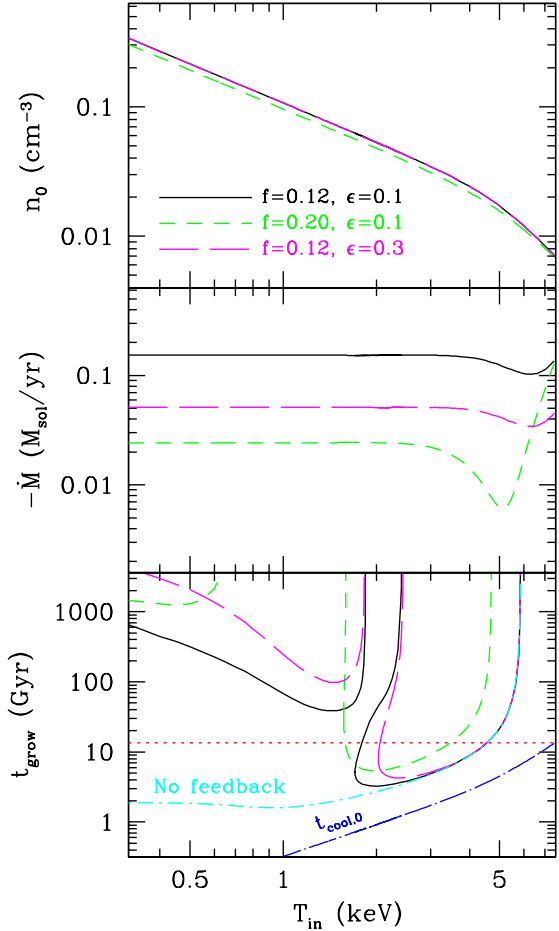}
\caption{Dependence of the cluster stability on the background profile for the cluster A1795. For a fixed value of $f$ and $\epsilon$, the corresponding central electron number density (\textit{upper panel}), steady-state mass accretion rate (\textit{middle panel}), and the growth time of unstable modes (\textit{lower panel}) are plotted as a function of the central gas temperature $T_{\rm{in}}$. The \textit{dot short-dashed} line in the lower panel shows the growth time of the unstable mode in steady-state models with $f=0.12$, $\epsilon=0.1$ and no feedback mechanism for AGN heating (i.e., $\Delta \mathcal{H}_{\rm{feed}}$ in equation \ref{stab4} is taken to be zero), while the \textit{dot long-dashed} line stands for the central gas cooling time in models with $f=0.12$, $\epsilon=0.1$.}
 \label{plot8}
 \end{figure}
 
\begin{figure}
\epsscale{1.0}
\plotone{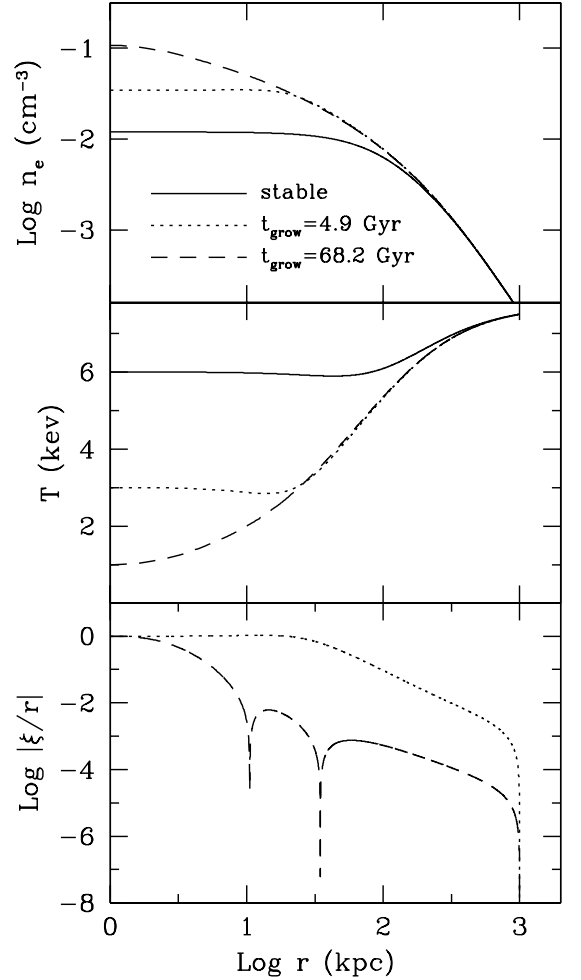}
\caption{Radial profiles of electron number density (\textit{upper panel}), temperature (\textit{middle panel}) and the eigenfunction of the radial unstable mode (\textit{lower panel}) in three typical steady-state models of the cluster Abell 1795 with $f=0.12$ and $\epsilon=0.1$.}
 \label{plot9}
 \end{figure}

\begin{figure}
\epsscale{1.0}
\plotone{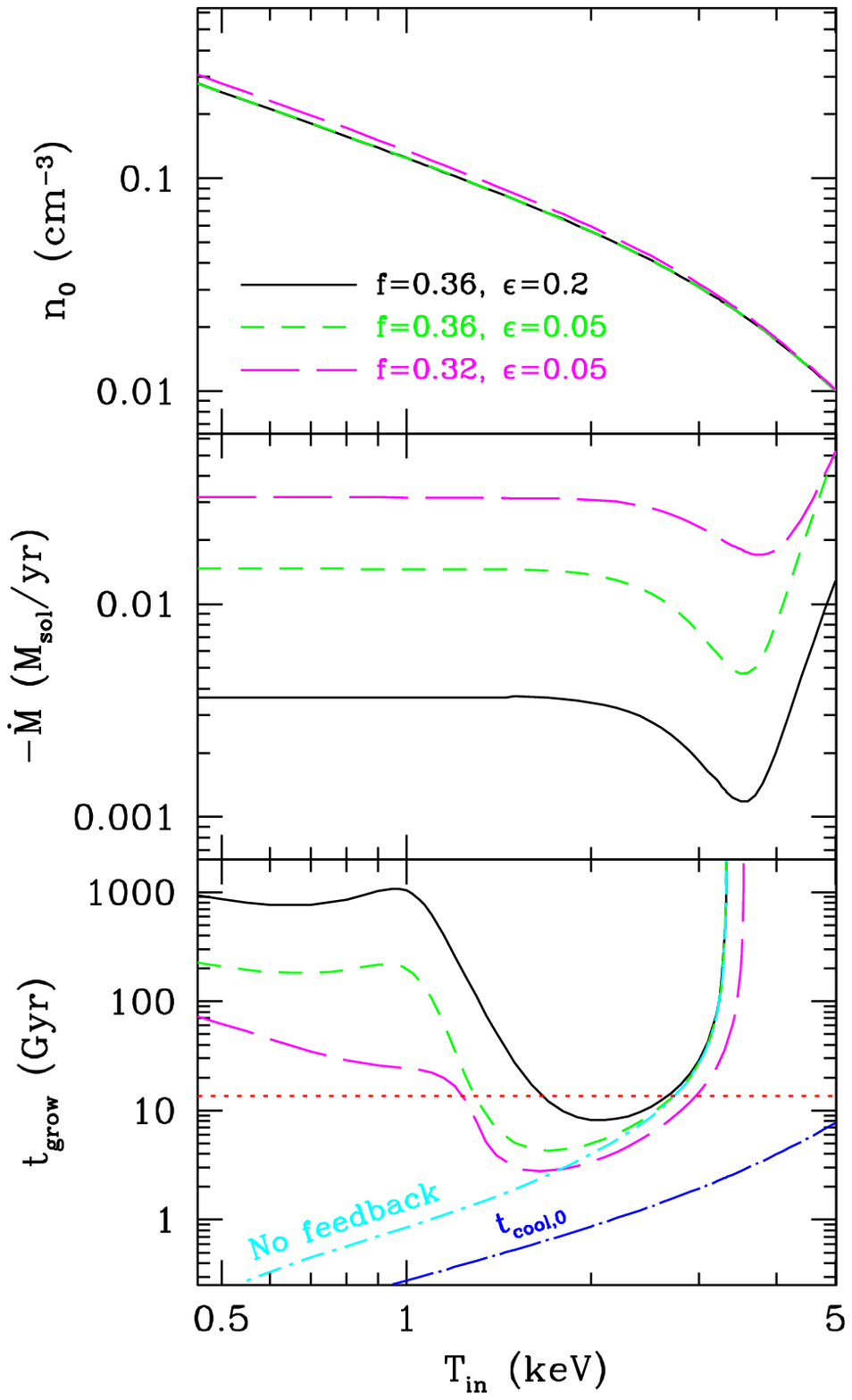}
\caption{Dependence of the cluster stability on the background profile for the cluster A2199. For a fixed value of $f$ and $\epsilon$, the corresponding central electron number density (\textit{upper panel}), steady-state mass accretion rate (\textit{middle panel}), and the growth time of unstable modes (\textit{lower panel}) are plotted as a function of the central gas temperature $T_{\rm{in}}$. The \textit{dotted} line in the lower panel stands for the Hubble time, while the \textit{dot short-dashed} line shows the growth time of the unstable mode in steady-state models with $f=0.36$, $\epsilon=0.2$ and no feedback mechanism for AGN heating, while the \textit{dot long-dashed} line stands for the central gas cooling time in models with $f=0.36$, $\epsilon=0.2$.}
 \label{plot10}
 \end{figure}
 
In this subsection, we will study the dependence of the cluster stability on the background steady-state ICM profiles. We adopt the clusters Abell 1795 and Abell 2199 as our fiducial clusters. Since the gas cooling time at the outer boundary is much longer than the Hubble time, we consider steady-state cluster models with fixed values of $T_{\rm{out}}$ and $n_{\rm{out}}$, which are chosen to be the same as those in the second model listed in Table \ref{table1} for each cluster. Of the three inner boundary conditions in equation (\ref{boundcond}), we choose two (including $r_{\rm in}^{2} F(r_{\rm in})=0$ and a given value of $T_{\rm in}$ for a specific model). We have four boundary conditions for three first-order ordinary differential equations (\S~\ref{section:stst}). Thus, the value of $\dot{M}$ can be solved as the eigenvalue for a specific steady-state model with a given value of $f$ and $\epsilon$ (which are deemed to be fixed by the physics of thermal conduction and black hole accretion respectively). The corresponding central electron number density for each steady-state model (represented by the varying value of $T_{\rm in}$) is shown in the {\it upper} panels of Figure \ref{plot8} and \ref{plot10}. In addition, it is possible to vary the inner boundary $r_{{\rm{in}}}$. Varying $r_{\rm in}$ at fixed $T_{\rm in}$ is basically denegerate with varying $T_{\rm in}$ at fixed $r_{\rm in}$, and thus we do not explore this additional dependence. 

We first consider the models of the cluster A1795 with $f=0.12$ and $\epsilon=0.1$ (i.e., variations of model A2). The dependence of the cluster stability on the background profile (represented by $T_{\rm{in}}$) is plotted in Figure \ref{plot8} (solid line), which clearly shows that the model with either a relatively flat ($T_{\rm{in}}>4.5$ keV) or steep ($T_{\rm{in}}<1.7$ keV) temperature profile is (effectively) stable, while thermal instability can develop at intermediate temperatures. Figure \ref{plot9} shows radial profiles of electron number density, temperature and the eigenfunction ($\xi/r$) of the radial unstable mode for three typical steady state models with $T_{\rm{in}}=6$ keV, $3$ keV, and $1$ keV respectively. 

\subsubsection{Cool core versus non-cool core clusters}

X-ray observations also suggest that clusters can be subdivided into two distinct categories according to the presence or absence of a cool core (e.g., Peres et al. 1998, Bauer et al. 2005, Sanderson et al. 2006, Chen et al. 2007).
In this section we discuss how our model may account for this effect.

The \textit{dot-dashed} line in the lower panel of Figure \ref{plot8} shows the growth time of the unstable mode in steady-state models with $f=0.12$, $\epsilon=0.1$ and no feedback mechanism for AGN heating (i.e., $\Delta \mathcal{H}_{\rm{feed}}$ in equation \ref{stab4} is taken to be zero). Without the feedback mechanism, the instability growth time in cool core clusters with relatively steep temperature profiles is very short ($\sim 2$ Gyr), suggesting that the feedback mechanism plays a key role in stabilizing thermal instability in these clusters. As the central gas temperature increases and the central gas density decreases, the stabilizing effect of the feedback mechanism becomes smaller, which is reasonable since the perturbation of the central mass accretion rate (note that $L_{\rm{agn}} \propto \dot{M}_{\rm{in}}$) scales as the central gas density (equation \ref{pertfeed}) and the importance of the feedback mechanism in the energy perturbation equation (\ref{gstab3}) may be expressed as $\Delta \mathcal{H}_{\rm{feed}}/(P\sigma \nabla \cdot \mbox{\boldmath $\xi$})$, which (using equations (\ref{agnheat}, \ref{hydro4}, \ref{delta_H_feed_simple}) and $dP/dr \sim \rho g$) scales as $T^{-2}$ at any specific radius. Colder, denser (i.e. low entropy) gas has a steeper pressure gradient in a fixed potential well, which increases the volumetric rate at which cavities perform $pdV$ work as they rise. 

On the other hand, although the stabilizing effect of the feedback mechanism becomes negligible for non-cool core clusters with relatively flat temperature profiles, our stability analysis surprisingly shows that these cluster models are also (effectively) stable. As shown in \S \ref{localrad}, the diffusive AGN heating ($\mathcal{H} \propto \partial P/\partial r$) only increases the growth time of local thermal instability, while thermal conduction may completely suppress local perturbations with wavelength less than $\lambda_{\rm{Field}}$, which increases as the central gas temperature increases and which may then be greater than the cooling radius.
Thus, we expect that thermal conduction alone may completely suppress thermal instability in these non-CC clusters.
This is confirmed by our stability analysis: we constructed NCC models with only conduction and found them to be stable. 

The short and long dashed lines in Figure \ref{plot8} show the models with a higher thermal conductivity and AGN feedback efficiency respectively. Obviously, a higher level of thermal conduction increases the stability of the cluster. However, a higher level of AGN feedback efficiency only increases the stability of CC cluster models, but has a negligible stabilizing effect on non-CC cluster models. Thus, AGN feedback heating is not required in NCC clusters. On the other hand, for CC clusters with lower central temperatures, AGN heating is generally required: otherwise one is either unable to build an equilibrium profile with a physically plausible level of thermal conduction $f < 1$, or else the CC cluster is globally unstable on fairly short timescales. We did the same calculations for the cluster A2052 and A2597, and found similar results.

We stress that, except for models with very high values of $f$ and $\epsilon$ (which may always be stable, as seen from the trend of different lines shown in Figure \ref{plot8} and \ref{plot10}), the clusters are usually stable either when the central temperature is high
(non-CC) or when the central temperature is sufficiently low. Figure \ref{plot8} (and the analogous Figure \ref{plot10} for A2199) suggest that the 
intermediate values correspond to globally unstable solutions.
This is consistent with 1D hydrodynamic simulations of RB02 and \citet{2007arXiv0706.1274G}: if conductivity is large, the cluster relaxes to stable NCC states (see Figure 1 and 2 of \citealt{2007arXiv0706.1274G}); if $f$ is small (i.e., conduction couldn't offset the cooling), the cluster first cools gradually through NCC states, and then quickly relaxes to stable CC states (see Fig. 1 in RB02 or Fig. 4 and 5 of \citealt{2007arXiv0706.1274G}).
We note that the radiative cooling times in non-CC clusters are generaly longer than in the CC ones and in some of the
non-CC clusters no heating is required to prevent thermal instability on a timescale shorter than the Hubble time
(note however that in the cases shown in Figures \ref{plot8} and \ref{plot10} (dot long-dashed lines), 
the central cooling timescale is always shorter then the instability growth time).
Irrespectively of whether this is the case or not, our model naturally explains the dichotomy between CC clusters 
where AGN feedback signatures are observed (and where the AGN play the key role in establishing global stability) 
and the non-CC clusters, where conduction may play a role but where the AGN feedback is not readily observed. 
This trend for the non-CC (flat temperature) clusters to have no AGN as opposed to CC clusters with clear
AGN observational signatures is clearly seen in the recent 
data compiled by Dunn \& Fabian (2008). This is also consistent with the observation by Rafferty et al. (2008) who show
that the short central cooling time corresponds to younger AGN (i.e., shorter X-ray cavity ages).

\begin{figure}
\epsscale{1.0}
\plotone{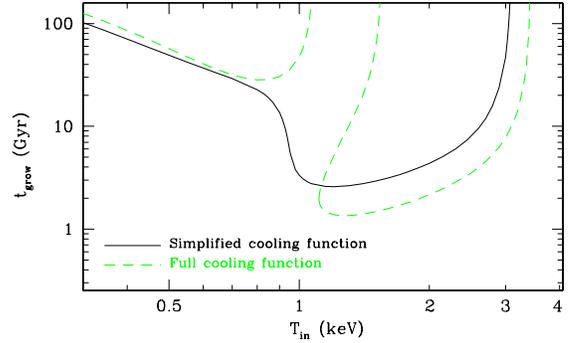}
\caption{The growth time of unstable modes for models of the cluster A2597 with $f=0.4$ and $\epsilon=0.05$, plotted as a function of the central gas temperature. Both the stable CC and NCC branches are seen in the calculations with either our simplified cooling function or a full cooling function (see text).}
\label{plot11}
\end{figure}

Metal line cooling may become important when the gas temperature is low. We checked our calculations with a full cooling function (equation 35 of \citealt{2007arXiv0706.1274G}, which is based on \citealt{1993ApJS...88..253S}), and did not find qualitative changes to our results. As an example, Figure \ref{plot11} shows the dependence of the cluster stability on the central gas temperature for the cluster A2597, which has the lowest $T_{\rm{in}}$ in our cluster sample.
With both free-free and metal line cooling included, the gas cooling rate increases and the dependence of $t_{\rm{grow}}$ on $T_{\rm{in}}$ becomes similar to that of the higher-temperature cluster A1795 (see Fig. \ref{plot8}), but both the stable CC and NCC branches still exist.

Through hydrodynamic numerical simulations, RB02 shows that the ICM heated by a combination of AGN feedback and thermal conduction usually relaxes to a stable quasi-steady state. Surprisingly, our stability analysis shows that a specific steady-state model may be globally unstable if the AGN feedback efficiency is lower than a critical value. This ``inconsistency" may be explained if the cluster with lower $\epsilon$ relaxes to a steady state with lower central gas temperatures, which may then be effectively stable, as clearly shown in this subsection. Recent numerical simulations by \citet{2007arXiv0706.1274G} indeed confirms that the cluster central regions in their cosmic-ray feedback models with lower $\epsilon$ cool to higher densities and lower temperatures in the final steady state (see Figure 7 of \citet{2007arXiv0706.1274G}).

\subsection{Global decaying modes}

\begin{figure}
\epsscale{1.0}
\plotone{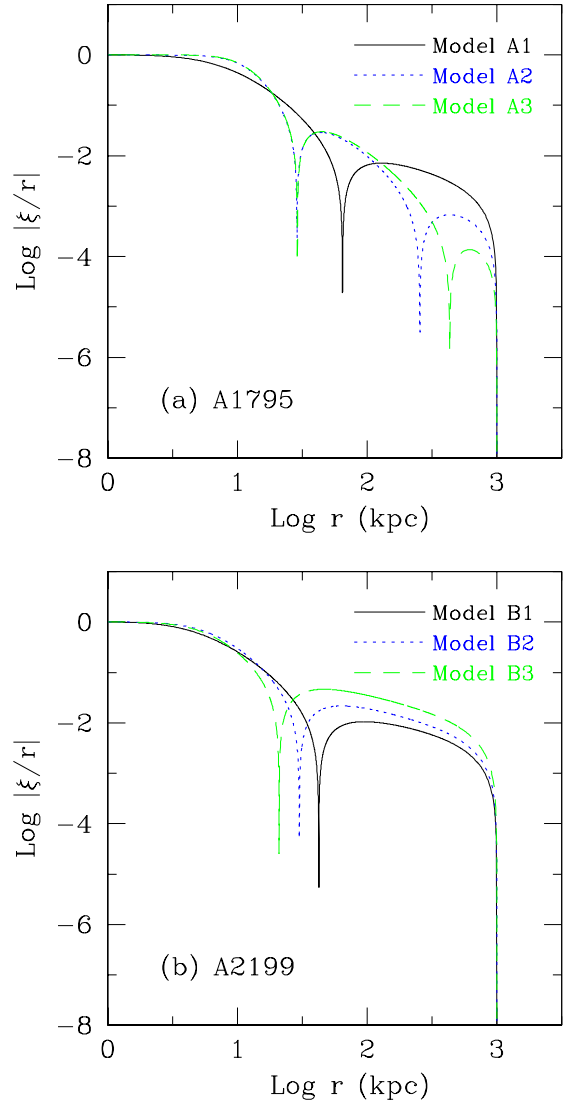}
\caption{Eigenfunctions of the slowest decaying modes ($\sigma<0$) in typical steady-state models for the cluster: (a) Abell 1795 and (b) Abell 2199. For each steady-state model, there exist decaying modes with larger decay rates, which are not shown in this figure.}
 \label{plot12}
 \end{figure}

We have also searched for global decaying modes with real and negative $\sigma$. For each steady-state model of the ICM, similar to KN03 (see Figure 4b of KN03), we found a series of decaying modes, within which there exists a slowest decaying mode with the smallest decay rate. Figure \ref{plot12} shows the eigenfunction ($\xi/r$) of the slowest decaying mode in typical steady-state models for the cluster Abell 1795 and Abell 2199. The decay rate of the most slowly decaying mode could potentially serve as an indicator of the ``attractor" solution toward which a cluster evolves after it has been reset by a merger, and might be worthy of further study. It may also be interesting to explicitly study the effect of AGN feedback on global non-radial modes and overstable modes, which we leave to future work (see \citealt{1987ApJ...319..632M}, \citealt{1989ApJ...341..611B} and KN03 for local analyses of these modes in the ICM with thermal conduction). 

\section{Summary and Discussion}
\label{section:conclusion}

Recent \textit{Chandra} and \textit{XMM-Newton} X-ray observations suggest that the hot intracluster gas is maintained by one or more heating sources at keV temperatures for a period at least comparable to the lifetime of galaxy clusters (\S~\ref{section:intro}). Since the emission lines expected from cooler gas are notoriously absent (e.g., \citet{2006PhR...427....1P}), the heating source (or sources) should also effectively suppress the thermal instability of the ICM. Although thermal conduction stabilizes thermal instability for perturbations with short wavelengths, the equilibrium ICM heated by thermal conduction alone is globally unstable and will evolve to a cooling catastrophe under global perturbations. On the other hand, using numerical hydrodynamic simulations, RB02 and \citet{2007arXiv0706.1274G} show that the ICM, which is initially in a state far from equilibrium and is heated by a combination of thermal conduction and AGN feedback, usually relaxes to a quasi-equilibrium steady state. Although these simulations adopt simplified 1D models for the elusive spatial distribution of AGN heating, they strongly suggest that the AGN feedback mechanism plays a key role in suppressing global thermal instability.

In this paper, we perform a detailed formal analysis of thermal instability in the ICM with both AGN feedback heating and thermal conduction. To build initial unperturbed states for stability analysis, we first construct steady-state cluster models, where the gas density and temperature profiles fit observations quite well and where the mass accretion rates are highly suppressed compared to those predicted by the standard cooling flow models (\S~\ref{section:stst}). Using the Lagrangian perturbation method, we then derive a set of differential equations that form an eigenvalue problem for global radial modes with the mode growth rate $\sigma$ as the eigenvalue. For pure conduction models of typical cool-core clusters, we find that the ICM has one unstable radial mode with the typical growth time $\sim 2-6$ Gyr (Table 2), which is consistent with the results of \citet{2003ApJ...596..889K}. However, for the hybrid models with both AGN heating and thermal conduction, global thermal instability is effectively reduced or even completely suppressed if the feedback efficiency $\epsilon$ is greater than a lower limit $\epsilon_{\rm{min}}$. Interestingly, if the AGN heating (equation \ref{agnheat}) is independent of the central mass accretion rate (i.e., no feedback mechanism for AGN heating), the ICM is still unstable (\S~\ref{section:globalstability}), which suggests that the feedback mechanism is essential to suppress thermal instability.

Assuming that the real intracluster gas is in a stable quasi-steady state, our global stability analysis thus suggests a minimum value $\epsilon_{\rm min}$ of the kinetic efficiency of AGN feedback, if it is to suppress a cooling flow. The value of $\epsilon_{\rm{min}}$ required in typical cool-core clusters is around $\sim 0.07-0.28$ (see Table 2), which is roughly consistent with the estimate of the jet production efficiency ($\sim 30\%$) for radio-loud AGNs by \citet{2007ApJ...658L...9H} and which is marginally consistent with recent observational estimations of $\epsilon \sim 0.01-0.1$ by \citet{2006MNRAS.372...21A} and \citet{2007MNRAS.381..589M}. Note that the value of $\epsilon_{\rm{min}}$ will be reduced if $\epsilon$ is an increasing function of the central mass accretion rate as suggested by recent observations (\S~\ref{section:deppara}).
Although the existence of $\epsilon_{\rm min}$ should be fairly robust, its exact value will also depend on the assumed form of AGN heating law. 

A related important issue in AGN feedback models is how and what fraction of the cooling gas at a distance of order $1$ kpc gets to the cluster center and finally fuels the AGN. This could be a very complex process due to the large range of distance scales involved. Our calculation extended from $\sim 1$Mpc in the cluster outer regions to $\sim 1$kpc at our innermost integration point. However, the black hole gravitational radius of influence $r_{\rm BH}$ is much further in:
\begin{equation}
r_{\rm BH} = \frac{G M_{\rm BH}}{\sigma^{2}} = 0.05 \, {\rm kpc} \left( \frac{M_{\rm BH}}{10^{9} \, {\rm M_{\odot}}} \right)\left( \frac{\sigma}{300 \, {\rm km\, s^{-1}}} \right)^{-2}  
\end{equation}
It is conceivable that if the flow is steady all the way down to the black hole, it transitions from a cooling flow to an adiabatic Bondi flow (see, for instance, \citealt{2000ApJ...528..236Q} and the discussion in \S~5 of \citealt{2007ApJ...671.1413C}), and there is tentative observational evidence that the Bondi formula may provide a reasonable approximation of the accretion process in X-ray luminous galaxies \citep{2006MNRAS.372...21A}. However, there are a host of potential complications: angular momentum, which would cause a torus or accretion disk to form instead, magnetic pressure and outflows \citep{2003ApJ...592..767P}, and local thermal instability resulting in the formation of stars and cold gas blobs \citep{2005ApJ...632..821P}, resulting in ``cold accretion" or the possibility that the black hole is fed by stellar winds. In this paper, in line with most numerical simulations in the literature (e.g., \citealt{2002ApJ...581..223R,2003ApJ...587..580B,2004ApJ...617..896H,2006ApJ...645...83V}) we have assumed that essentially all of the inflowing gas in our innermost boundary point eventually makes it to the black hole, over timescales long compared to the AGN duty cycle, but shorter than or comparable to the gas inflow timescale, $t_{\rm flow} \sim 10^{9}$yr. This behaviour is likely to be intermittent rather than steady: gas accumulates in an accretion disk/torus around the black hole, which eventually triggers an outburst and leads to gas consumption, etc. We assume that in steady state, the black hole will eventually consume all the gas which is supplied to it. In particular, in our model, while the black hole can exert {\it thermal} feedback on the cooling gas (through its heating activity), it exerts negligible {\it hydrodynamic} feedback (by consuming gas faster or more slowly than the supply rate, thus affecting pressure forces throughout the cooling flow region).  If indeed such a scenario applies, the results of our present paper should be recalculated in detail, since in that case $\dot{\rm M}$ cannot be freely varied, but is further constrained by the accretion law. Given the large uncertainties and difficulty of this calculation, we leave this to future work. 

In \S~\ref{section:depsteady}, we study the dependence of the cluster stability on the background steady-state profiles and find that the unstable cluster models with $\epsilon<\epsilon_{\rm{min}}$ may become effectively stable if the central gas temperature drops to much lower values. Numerical simulations by \citet{2007arXiv0706.1274G} indeed confirm that the cluster central regions in their cosmic-ray feedback models with lower $\epsilon$ usually cool to higher densities and lower temperatures in the final steady state. Thus, unlike the pure conduction models, where nonlinear evolution of global unstable modes lead to the cooling catastrophe, the ICM in our hybrid cluster models with AGN feedback included may always evolve to a quasi-steady state, which is effectively stable. On the other hand, we also show that thermal conduction will completely suppress thermal instability in non-CC clusters with relatively flat temperature profiles. Thus, the stability of the ICM favors two distinct categories of cluster steady state profiles: CC clusters stabilized mainly by AGN feedback and non-CC clusters stabilized by thermal conduction. Interestingly, recent X-ray observations also suggest that clusters can be subdivided into two distinct categories according to the presence or absence of a cool core 
(e.g., Peres et al. 1998, Bauer et al. 2005, Sanderson et al. 2006, Chen et al. 2007).

It is perhaps even possible that these two categories of clusters represent different stages of the same object. The importance of thermal conduction on global scales obviously depends on the large scale structure of the cluster magnetic fields. Recent calculations suggest that thermal conduction of heat into the cluster core can be self-limiting: in cases where the temperature decreases in the direction of gravity, a buoyancy stability sets in which re-orients a radial magnetic field to be largely transverse, shutting off conduction to the cluster center \citep{eliot08}. Non-linear simulations indicate the heat flux could be reduce to $\sim 1\%$ of the Spitzer value \citep{eliot_parrish08}. Thus, the following scenario could arise: as conductivity falls, gas cooling and mass inflow will increase, triggering AGN activity. The rising buoyant bubbles 
or the gas convective overturn mediated by cosmic rays (Chandran 2004)
may re-orient the magnetic field to be largely radial again, increasing thermal conduction and reducing mass inflow, shutting off the AGN until the heat flux driven buoyancy instability sets in once again. The cluster could therefore continuously cycle between cool-core (AGN heating dominated) and non cool-core (conduction dominated) states.  

Global linear stability analysis can clearly serve as a useful complement to simulations, both for the physical insight they can deliver and speed in exploring parameter space. Formally, a globally stable equilibrium state is a necessary but insufficient condition for a successful cluster model. A linear analysis fails for large non-linear perturbations (as a cluster would undergo, for instance, during mergers). Nonetheless, the linear stability analysis qualitatively reproduces the same features we have observed in 1D hydrodynamic simulations when we start the simulation from arbitrary initial conditions: conduction-only models suffer drastic cooling flows, while conduction + AGN heating models relax to a stable state with low mass inflow rates. For a given value of $\epsilon$ and $f$, a wide range of stable profiles are accessible. From a suite of linear stability analyses alone, it is not possible to predict the detailed final temperature and density profile the cluster relaxes to (which is somewhat sensitive to initial conditions), but it is possible to make general statements about stability and whether the cluster will relax to a cool core or non cool-core configuration. Finally, we have restricted our attention to a 1D radial analysis. While a 2D or 3D analysis would be more general, it would be much more complicated and unlikely to prove enlightening, at least for global modes. Non-radial modes may prove important in the case of local thermal instability, but the fastest growing unstable global mode---the possible cooling flow we wish to stem---is likely to respect spherical symmetry, unless the heating sources (such as the AGN jets) are severely anisotropic. Such models are beyond the scope of the present paper.   

\acknowledgements
We thank Omer Blaes, Ian Parrish, Mitch Begelman, Marcus Br{\"u}ggen, Paul Nulsen, Biman Nath for discussions. 
We also thank Mark Voit and Noam Soker for comments on the manuscript, and the anonymous referee for a very detailed and helpful report. FG and SPO acknowledge support by NASA grant NNG06GH95G. MR acknowledges support by {\it Chandra} theory grant TM8-9011X.

\bibliography{ms}

\label{lastpage}

\end{document}